\documentclass[twocolumn]{aastex631}
\usepackage{xspace}

\newcommand\CAS{CAS\xspace}
\usepackage{amsmath}
\received{June 29, 2023}
\revised{December 8, 2023}
\accepted{December 19, 2023}

\submitjournal{ApJ}

\shorttitle{High-redshift morphology}
\shortauthors{Tohill et al.}

\graphicspath{{./}{figures/}}

\begin{document}

\title{A Robust Study of High-Redshift Galaxies: Unsupervised Machine Learning for Characterising morphology with JWST up to z $\sim$ 8.}

\correspondingauthor{C. Tohill}
\email{clar-brid.tohill@nottingham.ac.uk}

\author[0000-0003-2527-0819]{C. Tohill}
\affiliation{University of Nottingham, School of Physics \& Astronomy, Nottingham, NG7 2RD, UK}
\author[0000-0001-7821-7195]{S. P. Bamford}
\affiliation{University of Nottingham, School of Physics \& Astronomy, Nottingham, NG7 2RD, UK}

\author[0000-0003-1949-7638]{C. J. Conselice}
\affiliation{Jodrell Bank Centre for Astrophysics, University of Manchester, Oxford Road, Manchester UK}

\author[0000-0002-8919-079X]{L. Ferreira}
\affiliation{University of Victoria, Astronomy Research Centre, Victoria, British Columbia, Canada}

\author[0000-0002-4130-636X.]{T. Harvey}
\affiliation{Jodrell Bank Centre for Astrophysics, University of Manchester, Oxford Road, Manchester UK}

\author[0000-0003-4875-6272]{N. Adams}
\affiliation{Jodrell Bank Centre for Astrophysics, University of Manchester, Oxford Road, Manchester UK}

\author[0000-0003-0519-9445]{D. Austin}
\affiliation{Jodrell Bank Centre for Astrophysics, University of Manchester, Oxford Road, Manchester UK}

\begin{abstract}
Galaxy morphologies provide valuable insights into their formation processes, tracing the spatial distribution of ongoing star formation and encoding signatures of dynamical interactions. While such information has been extensively investigated at low redshift, it is crucial to develop a robust system for characterising galaxy morphologies at earlier cosmic epochs. Relying solely on nomenclature established for low-redshift galaxies risks introducing biases that hinder our understanding of this new regime.
In this paper, we employ variational auto-encoders to perform feature extraction on galaxies at z $>$ 2 using JWST/NIRCam data. Our sample comprises 6869 galaxies at z $>$ 2, including 255 galaxies z $>$ 5, which have been detected in both the CANDELS/HST fields and CEERS/JWST, ensuring reliable measurements of redshift, mass, and star formation rates. To address potential biases, we eliminate galaxy orientation and background sources prior to encoding the galaxy features, thereby constructing a physically meaningful feature space. We identify 11 distinct morphological classes that exhibit clear separation in various structural parameters, such as \CAS-$M_{20}$, S{\'e}rsic indices, specific star formation rates, and axis ratios. We observe a decline in the presence of spheroidal-type galaxies with increasing redshift, indicating a dominance of disk-like galaxies in the early universe. We demonstrate that conventional visual classification systems are inadequate for high-redshift morphology classification and advocate the need for a more detailed and refined classification scheme. Leveraging machine-extracted features, we propose a solution to this challenge and illustrate how our extracted clusters align with measured parameters, offering greater physical relevance compared to traditional methods.
\end{abstract}

\keywords{Astronomy, Galaxies --- 
Unsupervised Machine Learning --- Deep Learning --- High Redshift --- Morphology --- Classification}

\section{Galaxy Morphology} \label{sec:intro}
The morphology of a galaxy is a record of its formation history. It has been shown that the morphology of a galaxy traces the spatial distribution of ongoing star-formation and encodes the signatures of past and on-going dynamical interactions, which can give us an indication of how galaxies evolved throughout cosmic time \citep{Holmberg, Dressler, Kauffmann,Conselice_merger}. The Hubble classification scheme describes the morphologies of galaxies observed in the local universe. This classification scheme not only describes the visual appearance of the galaxy, but it has been shown that morphological type correlates strongly with many intrinsic properties such as the star formation rate (SFR), age, number of past merging events etc \citep{Sandage_hubble, lotz_merger}. However, while this system has recently been shown to describe galaxies up to very high redshifts $z \geq 5$ \citep{leo_hubble,Kartaltepe_jwst,jacobs_jwst,Huertas-company_jwst}, the number of irregular galaxies increases rapidly with redshift \citep{Abraham_irr, mortlock_candels, Leo_merger}, thus requiring a more detailed description.

Investigating galaxies at higher redshifts we observe much more peculiar and clumpy morphologies \citep{conselice_irr}. Due to the variety of galaxy types we observe it becomes challenging to create a  classification scheme for all galaxies akin to the Hubble tuning fork we apply at low redshifts. Relying on the nomenclature that has served us at low-redshift risks biasing our understanding of this new regime. 
For example, work carried out in the past by \cite{elmegreen_morph} classified high redshift galaxies in the Hubble Ultra Deep Field into 6 main groups; chain, clump cluster, double clump, tadpole, spiral and elliptical. These groups were determined by eye to match matched previous classifications by others \citep{Cowie_morph, vandenbergh_morph}. While many galaxies at high redshift fall into one of these categories, a more robust classification scheme is needed. 
The \CAS parameter system (concentration \citep{Bershady}, asymmetry \citep{schade}, and smoothness of a galaxy's light profile), defined in \cite{Conselice}, is one way this can be achieved, for example galaxies that have been classified as `Tadpole' galaxies tend to have high asymmetries while `clump cluster' galaxies tend to have low concentration values. Similarly the Gini-$M_{20}$ (G, $M_{20}$) non-parametric measurement system introduced by \cite{Lotz-gini} showed that combining the \CAS parameters with G and $M_{20}$ was a more effective approach to classifying different morphologies.
While these parameters can aid in distinguishing between certain classifications, it remains unclear if all high-redshift galaxies neatly fit into these morphological categories and how these relate to quantitative structure and other physical properties. Therefore, it is crucial to develop an efficient and robust method that groups galaxies based on their intrinsic features, without imposing our own biases on the classification classes and criteria

A classification scheme for early galaxies would allow us to probe deeper into the intrinsic properties of these galaxies and will help to better understand their evolution. We know that the star formation rate in the universe peaked at around $z \sim 2$ \citep{Madau_sf}, meaning that galaxies would have had more star forming regions, leading to more clumpy morphologies. 
We also know that the merger rates were higher in these earlier times \citep{Duncan,patton_merger}, meaning galaxies possess more disturbed morphologies with tidal disruptions and multiple cores etc. However, what is not understood is how these galaxies evolved into those we observe today. For example looking at galaxies, both star forming and quiescent, at a fixed stellar mass, we observe those at higher redshift to be more compact than their lower redshift counterparts \citep{Wilman}. How these evolved into what we see today could possibly be studied by investigating the evolution in morphological type with redshift. How to define morphological type is however not obvious at high redshift, which is the focus of this paper.

With the successful launch of the James Webb Space Telescope (JWST) we have access to the highest resolution imaging of these distant galaxies, allowing us to explore the high-redshift regime in the greatest depth and detail to date. This opens a window into better understanding the formation of the first galaxies and their evolution over cosmic time. There have already been a number of studies investigating the morphologies of these most distant objects \citep{leo_hubble,Huertas-company_jwst,Guo_bars_jwst}. While these studies focus on the morphology of these distant galaxies, they still characterise them using nomenclature used at low-redshift, investigating fractions of spheroid, disk and irregular type morphologies. While some galaxies at high-redshift fall into one of these groups, it is not known if this is applicable to all galaxies we observe in the distant universe. Nor is it known what features are important in concluding what morphological group a galaxy falls into. 

The problem that we investigate is how do we robustly classify these distant galaxies into self-similar types? How do we determine what features of a galaxies morphology are most important in its characterisation? Previous attempts to solve this issue involve citizen science projects such as Galaxy Zoo \citep{Willett}. These aim to amass a large number of visual classifications by asking the public to answer a number of questions about a galaxies shape, color etc. \citep{Bamford, Cardamone,Schawinski}. However, there are a finite number of questions and features that each classifier is able to select. This functions well for galaxies in the local universe and up to z$\sim$1, where the majority of galaxies fall into broad classifications of spiral, elliptical and irregular, however, this breaks down at the higher redshifts where the majority of galaxies lie in this irregular group. In order to better classify galaxies at high-redshift using this method there needs to be new questions and features available for each classifier to choose. The issue is these features are unknown, as there is no robust classification scheme in the distant universe. There is solution to this problem, and one that has become popular in recent years – machine learning.
Work has been carried out by \cite{Walmsley_zoo} that combines these visual classifications from Galaxy Zoo with machine learning, allowing for many more classifications, and also enabling researchers to locate anomalies within their datasets \citep{Walmsley_anomoly}. While these techniques have proven very successful, they still require labels to initially train the networks.  
By using unsupervised machine learning we can remove the need for any classification or labels. 

In this work we utilise an unsupervised deep learning algorithm to extract the most dominant morphological features for distant JWST galaxies and separate them into self-similar types, allowing for a new broad classification system. The intrinsic properties of the galaxies within each group are investigated with redshift, mass, and star formation rates to provide new insights into the evolution of galaxy structure since $z < 8.$

This paper is organised as follows. In \S\ref{sec:data} we introduce the imaging data and survey used in this work, along with our selection criteria. In \S\ref{sec:method} we detail the various architectures we explore in this work as well as the clustering algorithm used. We discuss our data standardisation process in \S\ref{sec:standardisation} and the optimisation of our networks in \S\ref{sec:opt}. The clustering algorithm used in this work is detailed in \S\ref{sec:clustering}. The resulting optimised network and results are included in \S\ref{sec:results}, which also includes information about the extracted morphologies and clusters. We conclude with a brief summary of our main results in \S\ref{sec:summary}.

\section{Data} \label{sec:data}

\subsection{JWST data}
All of the images used in this project are from the Cosmic Evolution Early Release Science Survey (CEERS; PI: Finkelstein, ID=1345, \cite{Finkelstein}) pubic release fields \citep{ceers_overview}, imaged with the NIRCam instrument on the James Webb Space Telescope (JWST). NIRCam offers wavelength coverage from $0.6-2.3\mu m$ with a resolution of $0.031^{\prime\prime}$/pixel, and from $2.3-5\mu m$ with a resolution of $0.063^{\prime\prime}$/pixel.

\begin{figure}[ht]
    \centering
    \includegraphics[width=\linewidth]{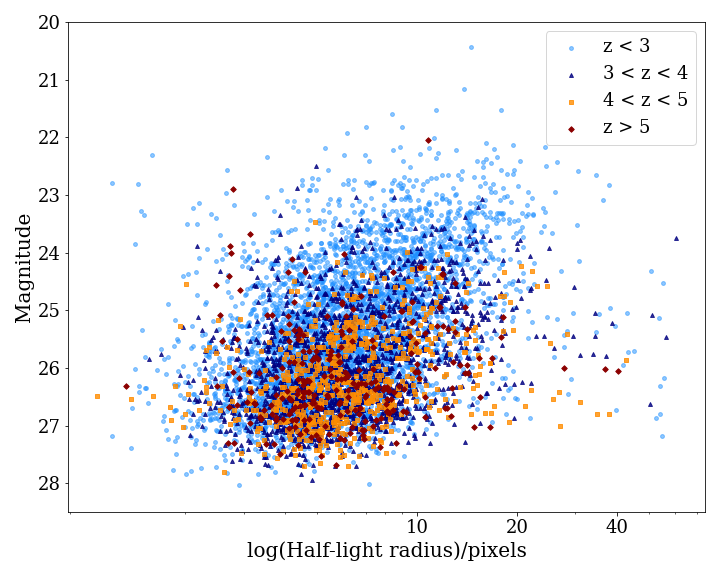}
    \caption{Distribution of the apparent $H_{160}$ magnitude of our galaxy sample vs. half-light radius in pixels. The resolution from NIRCam is $0.03^{\prime\prime}$ and $0.063^{\prime\prime}$ per pixel for $0.6-2.3\mu m $ and $2.3-5\mu m$ respectively. Our sample is measured from the JWST CEERS imaging.} 
    \label{fig:mag_dist}
\end{figure}

\begin{figure}[ht]
    \centering
    \includegraphics[width=\linewidth]{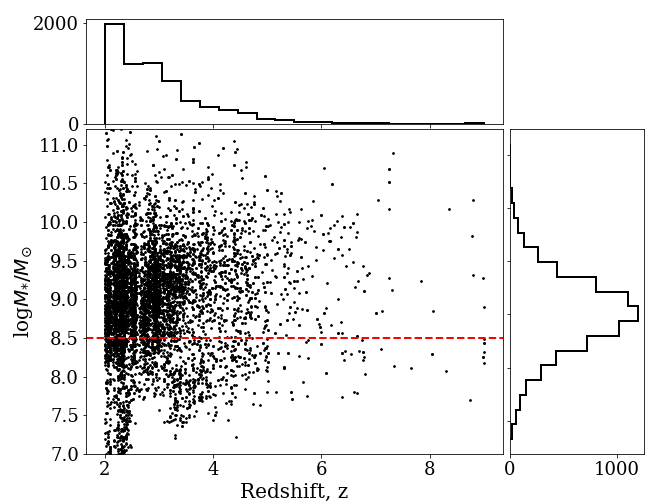}
    \caption{Distribution of the stellar masses of galaxies within our matched sample vs redshift. Red line shows mass cut at $10^{8.5}M_{\odot}$. Galaxies above this mass are utilised in this work which is above the completeness limit at all redshifts.} 
    \label{fig:mass_cut}
\end{figure}

The data has been reduced using the pipeline mentioned in \cite{leo_hubble}. This is a modified version of the JWST official pipeline 1.6.2, see \cite{leo_hubble, Adams_pipeline} for more detail.  
We select galaxies that overlap with the Cosmic Assembly Near-IR Deep Extragalactic Legacy Survey (CANDELS; \cite{CANDELS, Koekemoer_CANDELS}) for this work. In total we have $\sim$ 10,000 galaxies from CEERS that overlap with sources in the CANDELS field across all redshifts. The reason for matching our JWST sample to HST imaging is so that we can use the reliable and well tested redshifts, photometry, mass and star formation rate (SFR) measurements that have been derived and utilised in previous works \citep{Duncan, Whitney}. Known AGN have also been removed. In total we select 6869 galaxies with $z > 2$ that have a match in the CANDELS survey. The apparent magnitude-size distribution of our sample is shown in Fig.\ref{fig:mag_dist}.  While we will be using the redshift, SFR, mass and other measurements from the original CANDELS galaxies, we re-measure the non-parametric morphology with \textsc{Morfometryka} to have updated and more accurate CAS, S{\'e}rsic, Gini-M20 etc values. We expect these to change from the values measured off the HST images due to the increase in resolution and signal-to-noise of the data. It should be noted that these measurements were performed on the original galaxy stamps from JWST before any standardisation, discussed in \S\ref{sec:standardisation}, was applied. As we want to probe the rest frame optical wavelength for all of the galaxies in our sample, we use imaging from the F150W, F200W, F277W, F356W, F410M, F444W bands and match to the redshift of each galaxy, see Table \ref{tab:band_z} for redshift cuts.

\begin{table}[ht]
    \centering
    \begin{tabular}{cccc}
        \hline
             \textbf{Redshift} &   \textbf{Band} & \textbf{Total} & $\lambda_{pivot}$ ($\mu$m)\\
             \hline
             $ 2.0 \le z < 3.0$ & F150W & 4186 & 1.501 \\ 
             $ 3.0 \le z < 4.0$ & F200W & 1672 & 1.990 \\ 
             $ 4.0 \le z < 5.0$ & F277W & 644 & 2.786 \\ 
             $ 5.0 \le z < 6.0$ & F356W & 142 & 3.563 \\ 
             $ 6.0 \le z < 7.0$ & F410M  & 71 & 4.092\\
             $ z \geq 7.0 $ & F444W  & 42 &4.421 \\
    \end{tabular}
    \caption{Summary of the bands utilised from JWST NirCam with the associated redshift that aligns with the optical rest-frame.}
    \label{tab:band_z}
\end{table}{}
In order to ensure that the conclusions we draw in this work are representative of the sample we make a mass cut at $M_{*} > 10^{8.5}M_{\odot}$ to ensure we are complete at the highest redshifts. This can be seen in Fig.\ref{fig:mass_cut}. Galaxies above this limit are used for the analysis in this work.

\section{Method} \label{sec:method}
\begin{figure*}[htb!]
    \centering
    \includegraphics[width=\linewidth]{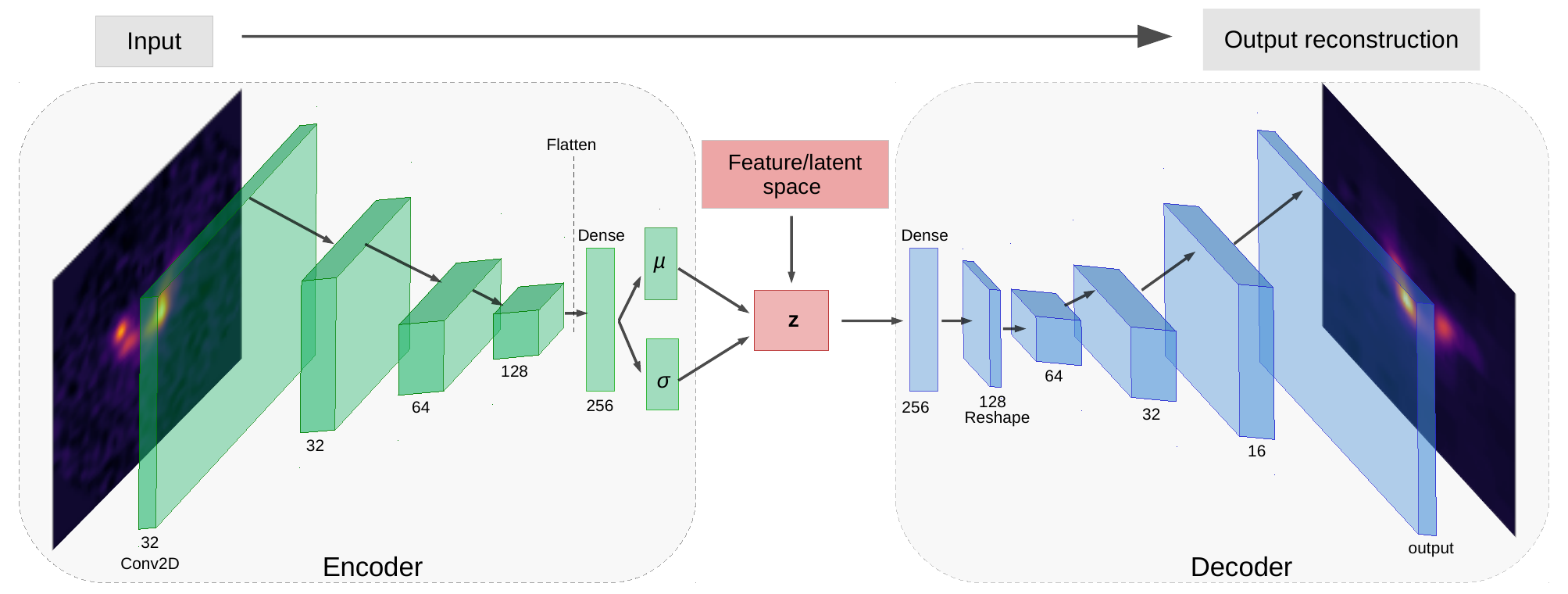}
    \caption{ Architecture of our MMD-VAE network. On the left is the encoder, this takes the input data and compresses it into a smaller number of dimensions represented by the feature space (z). The decoder then samples from this feature space to create a reconstruction of the input. The aim of this work is to cluster this feature space to extract groups of galaxies with similar morphologies.}
    \label{fig:AE}
\end{figure*}
\subsection{Machine Learning}
In recent years Machine Learning (ML) has proven to be very successful in astronomy and has been applied to many different problems. 
ML has been utilised to predict many morphological parameters of galaxies from parametric measurements such as the S{\'e}rsic index \citep{Tuccillo, Li_sersic,Tortorelli_sersic}, to non-parametric structural measurements such as the $\CAS$ system \citep{CASNET}. Supervised ML has also been successfully applied to visual classifications such as mergers \citep{Leo_merger}, anomaly detction \citep{Walmsley_anomoly}, and Hubble type classifications \citep{Dieleman, Dominguez_SDSS, Sunny}. More recently \cite{Robertson_MLjwst} has used deep learning to uncover the abundance of `disky' objects at high-redshift with JWST.
While these studies have proven to be successful, they require prior knowledge of the data in order to have labels to train your network on. This comes with its own issues, firstly you need to amass enough labels to train your networks which is possible through citizen science projects such as Galaxy Zoo \citep{Lintott_gz}. However, as these classifications are made by humans they come with their own intrinsic biases due to the subjective nature of the classifier. When we use these labels to train ML networks we are propagating this bias forward into any future classifications as well. With the future of astronomy consisting of "Big Data" surveys, it will take hundreds of people years to classify all of the 1.5 billion resolved galaxies in the Euclid survey \citep{Euclid}. 
One solution to remove this bias and to improve the efficiency of these classifications is to move towards using unsupervised machine learning techniques. 

\subsection{Unsupervised Machine Learning} \label{sec:UML}
As the name suggests, unsupervised machine learning techniques require no labels to train but use only the data that you are interested in investigating as an input. For this reason unsupervised methods can be a more robust and unbiased method for data analysis. 
There have been studies in recent years that have applied unsupervised techniques to different problems in astronomy such as strong gravitational lenses \citep{Sunny_lens}, anomaly detection \citep{ outlier_galaxy, outliers_uml} and galaxy morphology \citep{Hocking, Martin, Sunny_UML}. These studies work by training a network to perform feature extraction on input data to recover the main features of your image. These features can then be explored and analysed, allowing you to perform different tasks such as grouping similar images together (a classification type analysis), finding outliers in the data (anomaly detection), looking for correlations between features and physical properties (morphology studies), etc. As the issue we are trying to address is a classification-type analysis we will need to group similar extracted features together.
To perform our feature extraction we explore the use of variational autoencoders (VAEs) \citep{Kingma_VAE} which we describe in detail below. 

\subsection{Variational Autoencoders}
In this work we utilise a type of autoencoder (AE) network. The main idea behind an autoencoding network is that of dimensionality reduction. Dimensionality reduction is the process by which the number of features needed to describe some data are reduced. 
An AE is composed of two main components, the encoder and the decoder (see Fig.\ref{fig:AE}). The encoder takes an input, which in this example is an image, and encodes the information into a lower dimensional representation of the data. This lower dimensional representation is stored in the latent space which is also referred to as the feature space. These will be used interchangeably throughout this work. The decoder then samples from this latent space to create a reconstruction of your input. The AE is trained to compress your input data whilst minimising the reconstruction loss between it and the output, decoded image.
One downfall to AEs is that they are prone to overfitting as there is no regularisation of this latent space. In order to combat this, we can use a VAE \citep{Kingma_VAE}. The main idea behind a VAE is that instead of encoding each extracted feature into one number, it is instead encoded as a distribution that the decoder can sample from to recreate your input. This forces the latent space to be smoother which can also allow for generative processes (i.e creating mock galaxy images). During the training of these networks the reconstruction loss is minimised the same way as before, however, there is an extra penalty on the network if the latent space diverges from a standard normal distribution. This is called the Kulback-Leibler (KL) divergence.  
While the traditional VAE has been used in many studies with success \citep{Thorne_vae,Xu_vae}, it has been shown that the extracted features can be entangled and difficult to separate into distinct, differentiable features. This is an issue for our work as we want to be able to compare the networks features to known and well established morphological properties e.g the concentration of light, close pairs, asymmetry etc. 
There are a number of variations to the VAE that aim to address this issue of entanglement. Two of the more success variations are the $\beta-\mathrm{VAE}$ \citep{Higgins}, and the MMD-VAE \citep{infovae}. While \cite{Locatello_disentangle} state that perfect disentangled representations are theoretically impossible, imperfect disentangling of features is an extremely useful concept and has been shown to be a very powerful tool in many disciplines \citep{Chen_disentangle,Higgins,Eastwood_disentangle}. We investigated both network architectures in our work to determine the optimal structure for our problem which we explain in detail below.

\subsubsection{$\beta-\mathrm{VAE}$}
The $\beta-\mathrm{VAE}$ architecture was first introduced by \cite{Higgins}. This type of VAE incorporates a weight on the KL loss to improve the disentanglement of features in the latent space. A value of $\beta$ = 1 represents the original VAE, a $\beta > 1$ forces a stronger constraint on the latent space to learn a more efficient latent representation of the data. The idea is that if there are some features of the data that are independent of each other then the network will be able to better disentangle them, leading to a more robust representation of the data. The loss is defined as;
\begin{equation}
    \mathcal{L}_{\mathrm{total}} = \mathcal{L}_{\mathrm{recon}} + \mathcal{L}_{\mathrm{KL}}
\end{equation}
where
\begin{equation}
    \mathcal{L}_{\mathrm{recon}} = \mathbb{E}_{q_{\phi}(z|x)}\log p_{\theta}(\hat{x}|z) 
\end{equation}
\begin{equation}
    \mathcal{L}_{\mathrm{KL}} = \beta D_{\mathrm{KL}}[q(z|x)||p(z)] = \beta \frac{1}{2} \sum_{j} 1 + (\log\sigma_{j})^2 - \mu_{j}^2 - \sigma_{j}^2
\end{equation}

Where $p_{\theta}(\hat{x}|z)$ is the likelihood of the data $\hat{x}$ given the latent space $z$ and $q_{\phi}(z|x)$ is the posterior distribution of your latent space. The aim is that the network will reduce the reconstruction loss between the input and the decoded data while at the same the KL divergence encourages the posterior to follow a distribution, normally a unit Gaussian. The reconstruction loss ($\mathcal{L}_{\mathrm{recon}}$) in this work is simply the mean square error (MSE) between the reconstructed image and the input data. The network will be penalised for diverging from either of these conditions, as is the case with the traditional VAE. In the $\beta-\mathrm{VAE}$ there is an additional adjustable hyperparameter $\beta$ that is introduced to the KL divergence term to balance this with the reconstruction loss.  
\cite{Higgins} showed that this additional hyperparameter is able to moderate the latent information  and force the network to learn a more efficient representation of the data that is also disentangled. 
As we are interested in comparing the extracted features to known morphological features this disentanglement is important, however while the KL divergence can be moderated it still penalises the latent space diverging from a unit Gaussian. While this is useful for generative purposes as it creates a smooth latent space to sample from, it may not be best suited for our proposes as we are interested in retrieving distinct sub-clusters within the feature space in order to obtain a robust separation of galaxy types. 
With this in mind we explore another variation of the VAE known as the MMD-VAE. 

\subsection{MMD-VAE}\label{sec:MMD}

The second network we investigate is the MMD-VAE. This type of network differs from the $\beta-\mathrm{VAE}$ in that it does not exploit the KL divergence but instead, finds the maximum mean discrepancy (MMD) \citep{Gretton_MMD} between your prior distribution $p(z)$ and the posterior $q(z)$. The MMD of two distributions is minimised when they are identical. Instead of comparing the overall distributions like the KL divergence, it samples from each distribution and compares the means of each sample. If these are very different it is unlikely that the two samples are from the same distribution. In order to sample from each distribution it uses the kernel trick. This allows non-linear data to be projected onto a higher dimensional space where it can be linearly divided by a plane. The MMD loss is thus defined as:
\begin{equation}
\begin{split}
\mathcal{L}_{MMD} = \mathbb{E}_{p(z),p(z')}[k(z,z^{\prime})] 
+ \mathbb{E}_{q(z),q(z^{\prime})}[k(z,z^{\prime})] \\
- 2\mathbb{E}_{p(z),q(z^{\prime})}[k(z,z^{\prime})]
\end{split}
\end{equation}
Where $k(z,z^{\prime}
)$ can be any universal function. The most common being a Gaussian kernel which we utilise in this work. 
The reconstruction loss is the same as before and so we end up with the total loss function in our MMD-VAE:
\begin{equation}
    \mathcal{L_{\mathrm{MMD-VAE}} = \mathcal{L}_{\mathrm{recon}} + \mathcal{L}_{\mathrm{MMD}}}
\end{equation}

The advantage MMD-VAE has over $\beta-\mathrm{VAE}$ is that it is not penalised for diverging from a Gaussian distribution as the loss is defined by the moments of the distributions and not the density. This is better suited to our problem as we ideally want a latent space that is easy to separate into clusters which is more achievable when the latent space is less compact/dense as it is when you have a high $\beta$ value in the $\beta-\mathrm{VAE}$ network. 

To fully compare both networks we optimise both and compare how the reconstruction loss varies between then for the same number of latent variables. This comparison is discussed in \S\ref{sec:opt}.

\begin{figure}
    \centering
    \includegraphics[width=\linewidth]{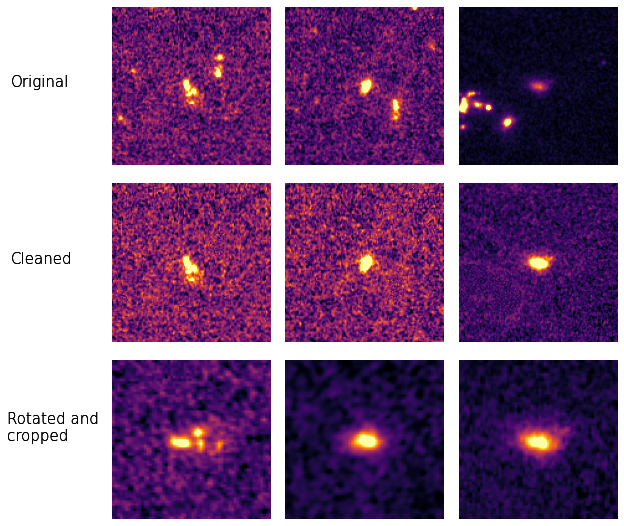}
    \caption{An example of our image standardisation process. \emph{Top:} Original CEERS JWST images. \emph{Middle:} This shows the galaxy images after they have been cleaned of background sources. \emph{Bottom:} Re-scaled and rotated images. }
    \label{fig:augment}
\end{figure}

\section{Data pre-processing}\label{sec:standardisation}
\subsection{Observational bias - rotation invariance} \label{sec:obs_bias}
One common issue that can arise with feature extraction is the fact that the network is trying to reproduce the input images with as few features as possible. This causes features such as shape, orientation, size and position to be encoded first as these will result in a smaller reconstruction loss than more finer details. These features however are not intrinsic to the galaxy and are in fact observational biases that we have imposed on the data simply because of our observation position on Earth. The dominance of these features has been well demonstrated in \cite{Astrovader}. In their work they show how almost half of their latent/information space encodes the orientation of their galaxies and the positions of background sources. While they were able to produce generated galaxy images with their network, they show one of the main downfalls of unsupervised machine learning techniques. 
Other authors such as \cite{Sunny_UML} address this rotation issue after feature extraction. During their clustering  of the extracted features, they use the rotation of each galaxy as a feature to define the clusters, thus avoiding galaxy orientation as a feature. This method was successful, however one downfall is that other structural features could be unintentionally excluded from the clustering method as some of their encoded information was used to encode this rotation. Addressing this rotation issue manually also means that this method is no longer totally unsupervised. 

In our work we want to address and remove these observational biases before trying to cluster our galaxies, thus allowing the feature space to be physically meaningful and without the risk of missing any other subtle features of the galaxies. 

\subsection{Image Standardisation}
To address these observational biases in our work we pre-process our data before training the network to prevent these features from being an issue, standardising our galaxy sample. 

An example of this is shown in Fig.\ref{fig:augment}. In the top row we have our original JWST images with the target galaxy in the center however, as you can see there are quite a number of background sources that as we have seen previously, a network will try to encode as a feature. We first remove these sources from our images using the galclean \citep{galclean} algorithm. This algorithm removes any non-central sources at a certain threshold above the background level. These masked areas are replaced with values sampled randomly from the background distribution to ensure they do not leave shapes which could be picked up by the network. The clean images can be seen in the middle row. The next issue we address is the orientation of the galaxies. As it has been shown in previous works this is one of the dominant features to a network and so we rotate all of our galaxies by their position angle to prevent this from becoming an issue. The position angle is measured on the JWST images to ensure that faint objects do not bias the standardisation process. The last feature we address is the apparent size of the galaxies. We re-scale all of our images to the average Petrosian radius ($\mathrm{R}_p$) of 15 pixels if they are smaller than this. This will allow the network to focus on the finer details of the images instead of wasting information encoding the size of the galaxies. We do not downscale our images as we do not want to lose any resolution as this will take away from the feature extraction and some finer features could be lost.
As our galaxies are all at high redshift we also crop our images to remove as much of the background as possible. An example of this can be seen in the bottom row of Fig.\ref{fig:augment}.
We then use these cleaned images to train our network.

\section{Model training and optimisation}\label{sec:opt}
Depending on the chosen network architecture, there will be a varying number of hyperparameters that need to be optimised. There are various methods to determine what set of hyperparameters will provide a suitable architecture for the problem being addressed. These range from the more basic random or grid search methods \citep{random_search}, to more advanced techniques such as random forest \citep{random_forrest}. These methods, whilst being used successfully in the past, are computationally expensive and can take a while to converge on an optimum value. A more efficient approach to optimising network architectures is Bayesian Optimisation \citep{Snoek}. Bayesian Optimisation builds upon previously evaluated models to create a probabilistic model which is built upon to more efficiently converge on an optimum solution. The hyperparameters within our network are as follows: the batch size fed to the network during training, the initial number of filters in the first layer of the network, the number of dense filters in the dense layer on the encoder, the optimiser used, the value of $\beta$ and $\lambda$ depending on the network being trained, and the number of latent dimensions used to encode our data. The latter we will address separately, as simply by increasing the number of latent dimensions the loss from our network will decrease, so to force the optimisation process to focus on the architecture of the network we keep the number of latent dimensions fixed at five. We chose this value as it is large enough to let the network encode the main features of each image so to have a reasonable loss to optimise the network on, but not too large that we risk encoding noise that would cause variations each time the network is trained. 
The optimum network hyperparameters are shown in Table.\ref{tab:opt_params}. For all networks trained the learning rate was reduced when the validation reconstruction loss had plateaued for a set number of epochs which is referred to as the `patience'. The patience for our learning rate was 20 epochs and if no improvement was seen after 50 epochs (the patience for the reconstruction loss) the training was stopped. All networks were allowed to run until there was no improvement seen in the validation reconstruction loss. 

\begin{table}
    \centering
    \begin{tabular}{|c|c|c|}
     \hline
     Hyperparameter & \multicolumn{2}{|c|}{Optimum value}  \\\cline{2-3}
      & MMD-VAE & $\beta-VAE$ \\
     \hline
     batch size & 32 & 32 \\
     fully-connected layer size & 256 & 128\\
     number of filters & 32 & 64\\
     optimization & Adam & Adamax\\
     $\beta$ &  N/A & 0.01  \\
     $\lambda$ & 10 &  N/A \\
     \hline
    \end{tabular}
    \caption{Summary of the hyperparameters selected by the Bayesian Optimization technique.}
    \label{tab:opt_params}
\end{table}

\begin{figure}
    \centering
    \includegraphics[width=\linewidth]{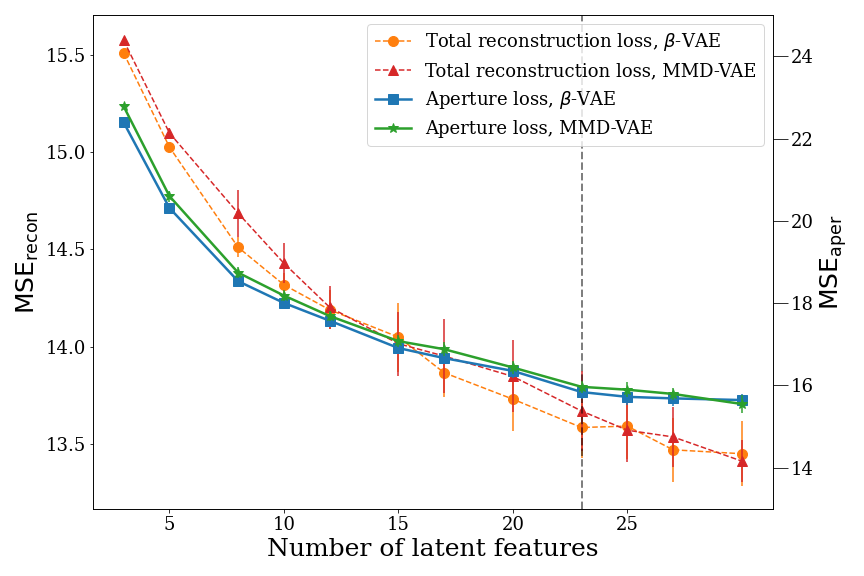}
    \caption{Variation of the reconstruction loss, both total and within an aperture, of the validation set from our networks vs the dimensionality of the latent space. Solid points show the average MSE for the last 50 epoches (our patience value, see text for details) and error bars show 1$\sigma$ of the variation. It can be seen that after 23 latent dimensions there is no improvement in aperture loss within the errors, whilst the total reconstruction loss continues to decrease.    
}
    \label{fig:latent_opt}
\end{figure}

\subsection{Dimensionality of latent space} \label{sec:dimension}
The main principle of a VAE is that of dimensionality reduction. Determining the optimum number of latent dimensions to encode the data into is another hyperparameter. A small number of latent features and you do not adequately capture the morphological information stored in each galaxy image. Too high, and you are left with a latent representation that is large and hard to interpret, making the idea of dimensionality reduction meaningless. To determine a good balance between these extremes we test 12 cases varying the dimensionality of the latent space from 3 to 30.
We use the reconstruction loss from our validation sample, which is simply the MSE between the reconstructed image from the decoder and the input image to determine the optimum number of latent features. We separate this loss into total reconstruction loss (i.e the whole image) $\mathcal{L}_{\mathrm{recon}}$, and the loss within an aperture of $R_p = 15$ pixels, $\mathcal{L}_{\mathrm{aper}}$, the average size of all our galaxy sample. The idea is that when the network has encoded the main features within each galaxy it will start to use information to encode the noise in the images. By exploring where the aperture loss plateaus we can select this to be the optimum number of latent dimensions to encode the morphological information of our sample of galaxies. The average variation in the loss for both networks is shown in Fig.\ref{fig:latent_opt}. The average loss is calculated on the validation set during the last 50 epochs of training each network which is the patience value for our training (i.e. when the training plateaus and there is no more improvement in the loss), and the error bars show the 1$\sigma$ variations in this loss.
Both networks perform similarly both at encoding the whole galaxy image and within $1R_p$ for each galaxy (see Appendix.\ref{fig:network_opt} for the training and test sets). For all latent dimensions investigated the total reconstruction loss continues to improve from a $\mathrm{MSE}_{\mathrm{recon}}$ of $\sim$15.5 to a $\mathrm{MSE}_{\mathrm{recon}}$ of $\sim$13.5, and continues to improve. However, it can be seen that after 23 latent dimensions there is no improvement within the aperture loss for either network. The $\mathrm{MSE}_{\mathrm{aper}}$ of the reconstructions within the aperture plateau at around $\sim$ 15.8 for both networks. This indicates that the network is utilising any extra dimensions to encode the noise in the image or background sources that might have been missed from the image cleaning process, which we also do not care about. This is also reflected in the larger error bars when we get to these number of latent dimensions. Thus for the analysis in this work we have a dimensionality of 23 for our feature space. 
We expected a smaller number of features would be needed to encode the main morphological structure of our sample as one of the main aims of this work is to minimise the latent space in order to ensure it is interpretable and physically meaningful. In earlier iterations of this work we utilised 5 latent dimensions to encode the data and found that, while the overall reconstructions were reasonable, the network tended to use individual latent features to encode a variety of physical features. This indicated that limiting the number of features to very low dimensionality limited both the networks reconstructions, and also limited us from mapping individual latent features to physical features. The fact that 23 latent dimensions are needed to well reconstruct our galaxy sample reflects the diversity of morphologies observed in the early universe at the resolution of JWST. This number of latent dimensions, whilst higher than first expected, is much lower than previous works that allow their feature space to get very large in order to achieve the best reconstructions possible, thus rendering their feature space un-interpretable and difficult to map to physical features.
For the rest of the analysis in this work we utilise the MMD-VAE network as both networks have a similar performance however for the reasons stated earlier in \S\ref{sec:MMD}, MMD-VAE has been shown to be better at disentangling features and puts less constraint on the distribution of the latent features, which is better for our problem.

\section{Clustering}\label{sec:clustering}
The aim of our work is to be able to separate our galaxies into different clusters based on their intrinsic features determined by our network. This is not a simple task, as we do not know the true number of clusters that exist in our data and so the question is, how do we determine how many different groups of galaxies exist in our data? 
To address this issue we explore a method known as hierarchical clustering, as no prior knowledge of how many `true' clusters there are within the data is required.

\begin{figure}[ht]
    \centering
    \includegraphics[width=\linewidth]{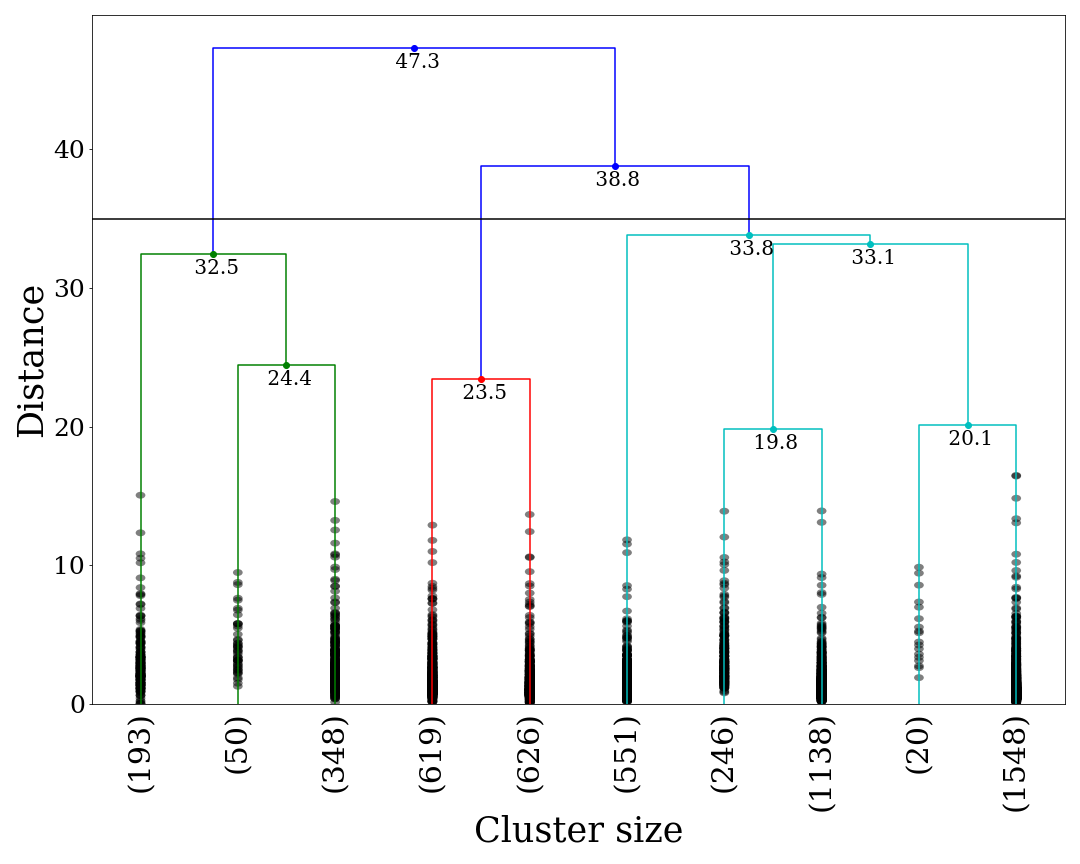}
    \caption{Dendrogram of the HC algorithm. The y axis shows the similarity or distance between the points while the x axis shows the number of data points in the smallest clusters. The horizontal black line shows the cut-off point in similarity that would result in 3 clusters.     
}
    \label{fig:HC_tree}
\end{figure}

\subsection{Hierarchical clustering}
Within this work we utilise the hierarchical clustering algorithm (HC) \citep{johnson_HC}.
We focus specifically on agglomerative HC which can be thought of as a `bottom up' clustering approach. This technique initially assumes that each point is its own cluster and will merge similar clusters together after every iteration until all data points are contained within one cluster. Clusters are merged or considered similar if they are close together in the feature space.
This method allows for uneven cluster sizes and also uneven shaped clusters. This allows more freedom in the latent space which is a feature of the MMD-VAE that we utilise in this work. An example of a HC dendrogram is shown in Fig.\ref{fig:HC_tree}. 
This method has been used in similar studies for classifying different morphological classes at low redshift \citep{Sunny_UML}.

In order to measure how similar two clusters are the Ward's linkage method is used. This method measures the variance within each cluster by means of the sum of squares within them, and aims to minimise this when grouping clusters together. The distance computed is thus the increase in the sum of squares when two clusters are merged. As this distance is minimised, the resulting clusters are created by grouping the closest points in our feature space together. 
In order to select clusters using this method, traditionally a single distance is used as a cut off point, selecting all clusters above this threshold. However, this is not applicable to galaxy morphology studies as different morphological types require more features to describe them than others, for example, spheroids require less information than mergers or spiral galaxies. In order to ensure that the clusters we extract are well separated and that we do not miss any due to a restrictive cut we explore splits down each branch in the HC tree until the split is due to the signal-to-noise of the sample or if the cluster has less than 50 samples, as this is $\sim$ 1\% of our sample and would not allow any meaningful analysis to be conducted in terms of redshift evolution or different SFRs. The extracted clusters from our trained feature space are discussed in \S\ref{sec:our_clusters}.

\section{Results}\label{sec:results}

\subsection{Image Reconstruction and feature extraction}

\begin{figure*}
    \centering
    \includegraphics[width=\linewidth]{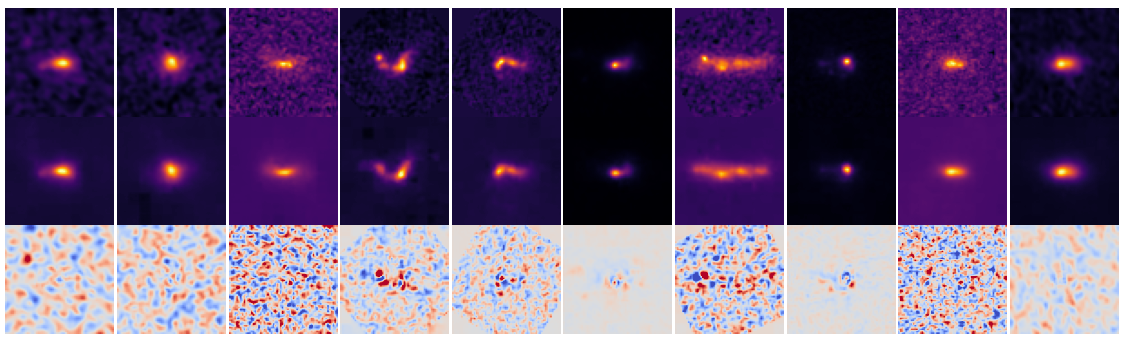}
    \caption{\emph{Top:} Input images to encoder. \emph{Middle:} Reconstructed images from the test set using 23 encoded features. \emph{Bottom:} Residuals showing how the reconstructions have encoded the galaxy light and not noise. It can be seen that the network is capturing the overall morphology of the galaxy sample well. Note that, whilst the main morphology of each galaxy is well reconstructed, some finer details are missed, such as the brightness of each clump. This is good for this work as we want to robustly classify different galaxy morphologies e.g. clumpy vs. not-clumpy morphology and not end up with many more groups split by small variations in the location of clumps as some previous works have been subject to. This will allow us to have a broader, more robust, separation of galaxy morphology. 
}
    \label{fig:recon}
\end{figure*}
We train our MMD-VAE network on a subset (80\%) of the images in our sample and use a validation set (10\%) to monitor the training to ensure the network is not over-fitting. We then compare the reconstruction loss between the validation data and a further independent test sample (10\%) to evaluate the performance of the network. We find a similar loss between these sub-samples indicating that the network architecture we have trained is robust.
An example of some reconstructions from the network can be seen in Fig.\ref{fig:recon}. The top row shows the input augmented JWST images, the middle shows the reconstructions from the network and the final row is the residuals between the input and the reconstructions. It can be seen that the network is able to encode the general morphology of each galaxy including shape, ellipticity, concentration and asymmetry of the light distribution, pairs and the clumpiness. This is a good sign as we do not see any orientation effects taking up any of the encoded information. We can also see the effect of information loss, our reconstructions are very smooth and have removed the noise from the images as well. 

Looking into the latent dimensions individually we can investigate which features the network is encoding. We have 23 latent dimensions in total that represent the feature space. In order to better understand the encoded space we investigate if there are any correlations between known galaxy parameters, both parametric and non-parametric, and individual latent features. We compute the Spearman's correlation co-efficient between each latent feature and our measured morphology to better understand what is being encoded. The highest ranked latent feature and the corresponding measurement can be seen in Table.\ref{tab:co-effs} (The full correlation matrix can be seen in Appendix.\ref{fig:spearman}).

\begin{table}[ht]
    \centering
    \begin{tabular}{ccc}
        \hline
             \textbf{Latent Feature} &   \textbf{Correlated feature} & \textbf{Spearmans rank} \\
             \hline
             Latent 1 & Axis ratio & 0.43  \\ 
             Latent 8 & Asymmetry & 0.49\\ 
             Latent 7 & S{\'e}rsic index & 0.38  \\ 
             Latent 8 & sSFR & 0.19 \\ 
             Latent 17 & Concentration  & 0.32 \\
    \end{tabular}
    \caption{The highest correlation value and corresponding galaxy property for each latent feature.}
    \label{tab:co-effs}
\end{table}{}

\begin{figure*}
    \centering
        \includegraphics[width= \columnwidth ]{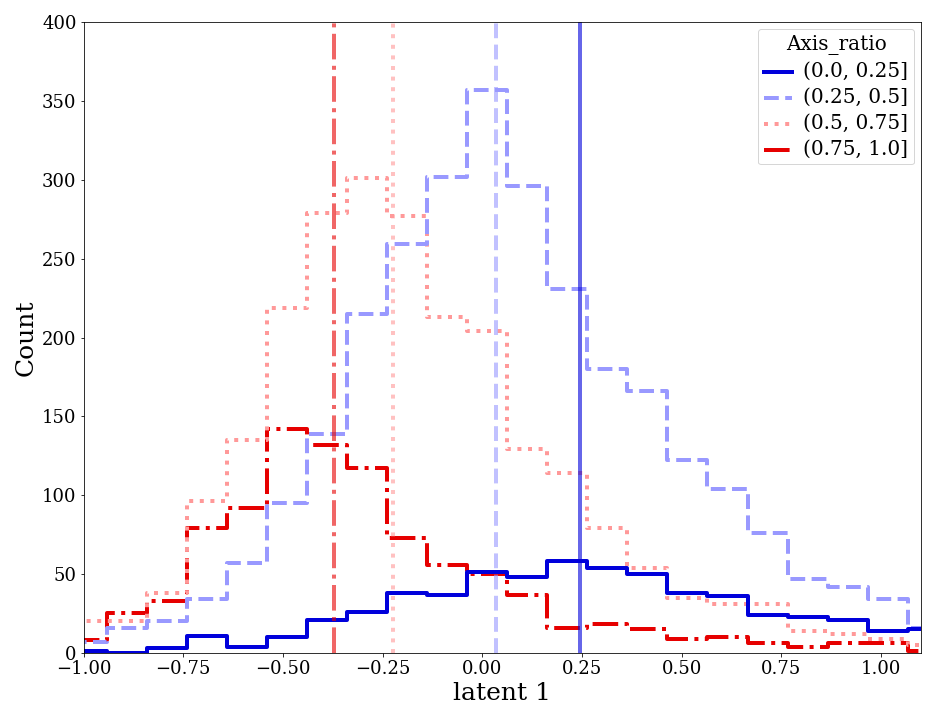}
        \includegraphics[width=\columnwidth]{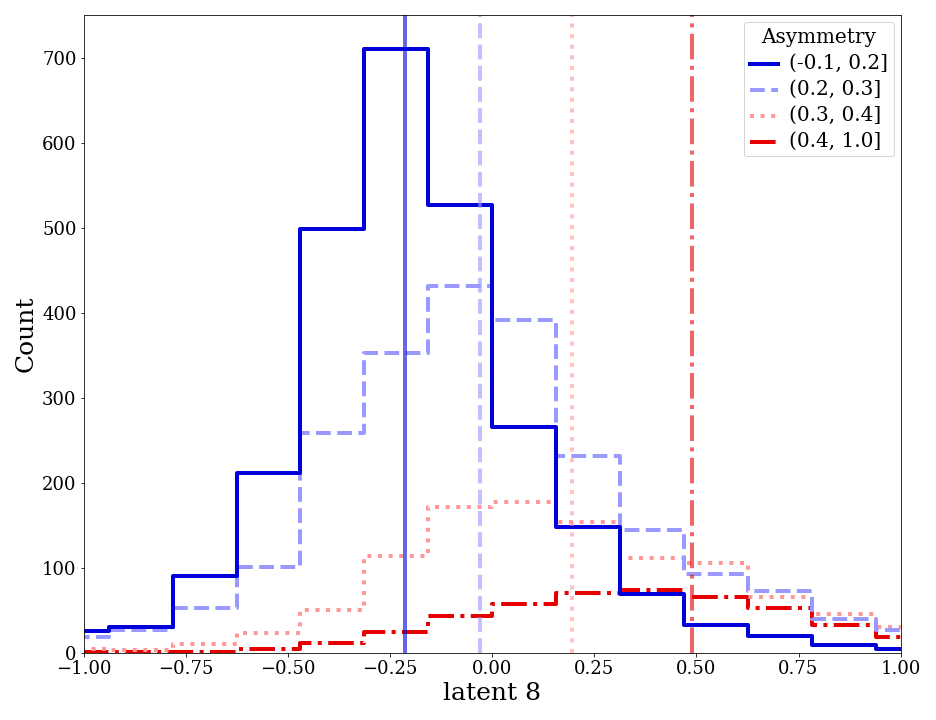}
    \\[\baselineskip]
        \includegraphics[width=\columnwidth]{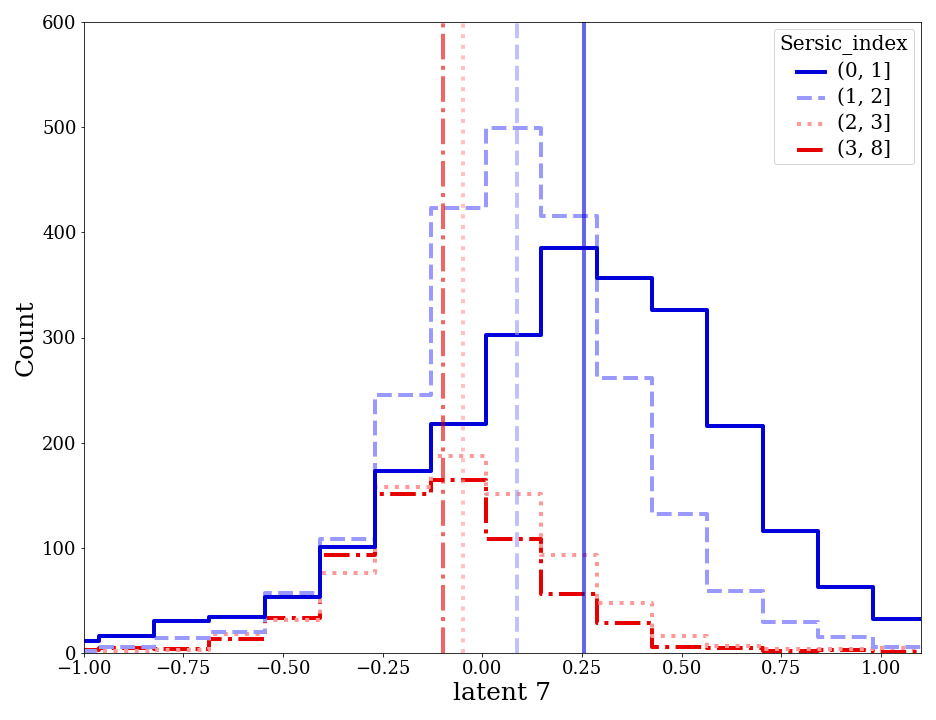}
        \includegraphics[width=\columnwidth]{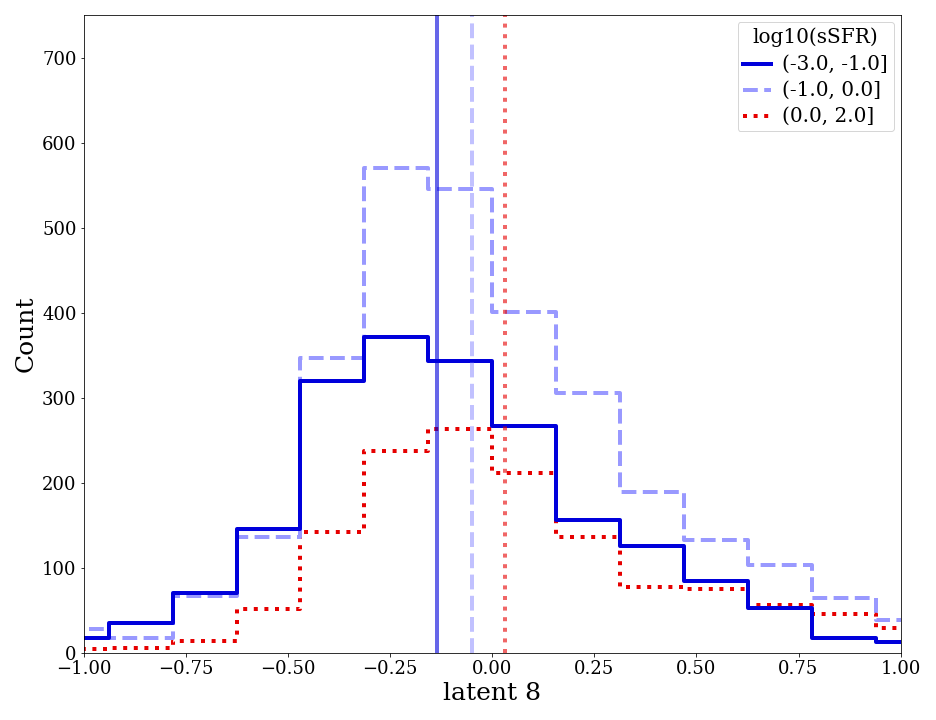}
    \\[\baselineskip]
        \includegraphics[width=\columnwidth]{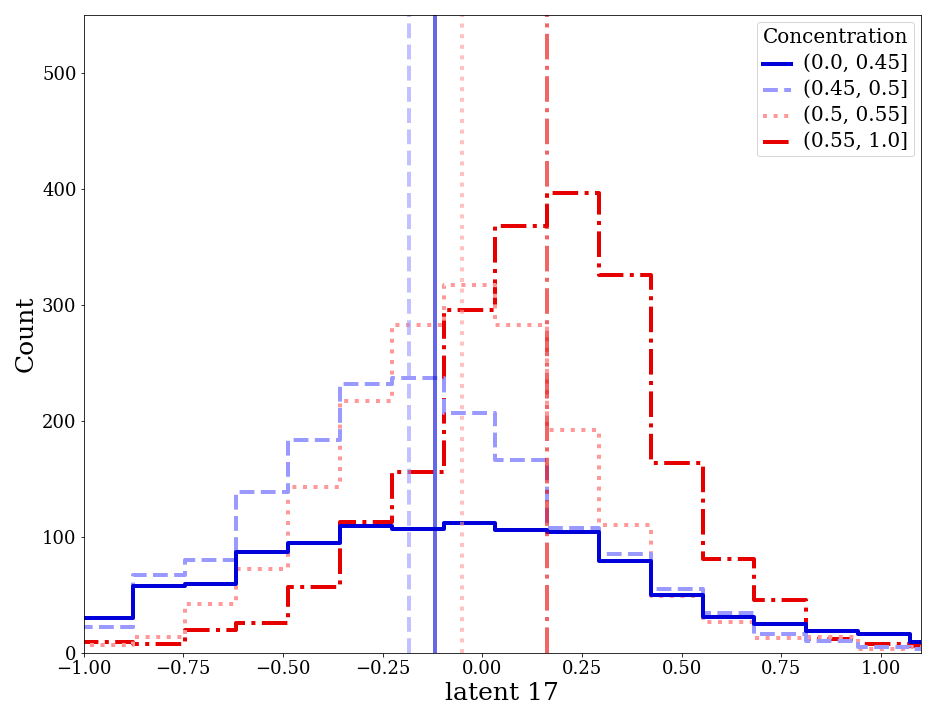}
    \caption{Correlation between features in the latent space and known measured properties for our galaxy sample. Each measured property is split into bins and plotted against the latent feature with which it had the largest Spearman's correlation coefficient. The median value for each selected range is also plotted. The ranges selected for each parameter can be seen in the legend. It can be seen that the features that the network is learning are physically meaningful and correlate with well know galaxy properties. } 
    \label{fig:latent_vs_feat}
\end{figure*}

In order to better visualise this correlation we plot each correlated feature vs the latent dimension with which it had the highest Spearman's co-efficient. We split each feature into bins to better visualise where different galaxy types lie along each feature. These can be seen in Fig.\ref{fig:latent_vs_feat}. Looking at the parametric and non-parametric properties we can see that they are well separated in each of these features which is important as it shows us that the network is learning features that are physically meaningful to our galaxy sample. We expect this from looking at the reconstructions from the network as they capture the overall morphology of our galaxies well. As a further demonstration, we show generated images from our learned latent space in Fig.\ref{fig:component_plot}. The generated images show a smooth transition in morphology along each latent feature as expected with a VAE. We include examples for the latent features that had the largest Spearman correlation coefficients with measured physical and structural properties as shown in Fig.\ref{fig:latent_vs_feat}.

We can see that while the sSFR does correlate with latent dimension 8 it is not as well separated as the other morphological measurements. This is to be expected as not all galaxies with high sSFRs will appear morphologically similar.
It has been observed that in the high-z universe high star-forming galaxies can be quite compact, and do not always resemble the classic star-forming clumpy morphology we think of traditionally. This makes it difficult to separate out one feature to describe how star formation affects galaxy morphology and makes visually identifying high star-forming galaxies biased and incomplete.  
By clustering galaxies based on a combination of all their extracted features we can avoid any pre-defined assumptions and uncover high star-forming galaxies with many different morphologies. We explore the morphology of high sSFR galaxies in \S\ref{sec:sSFR}.

\begin{figure*}[!htb]
    \centering
    \includegraphics[width=0.9\linewidth]{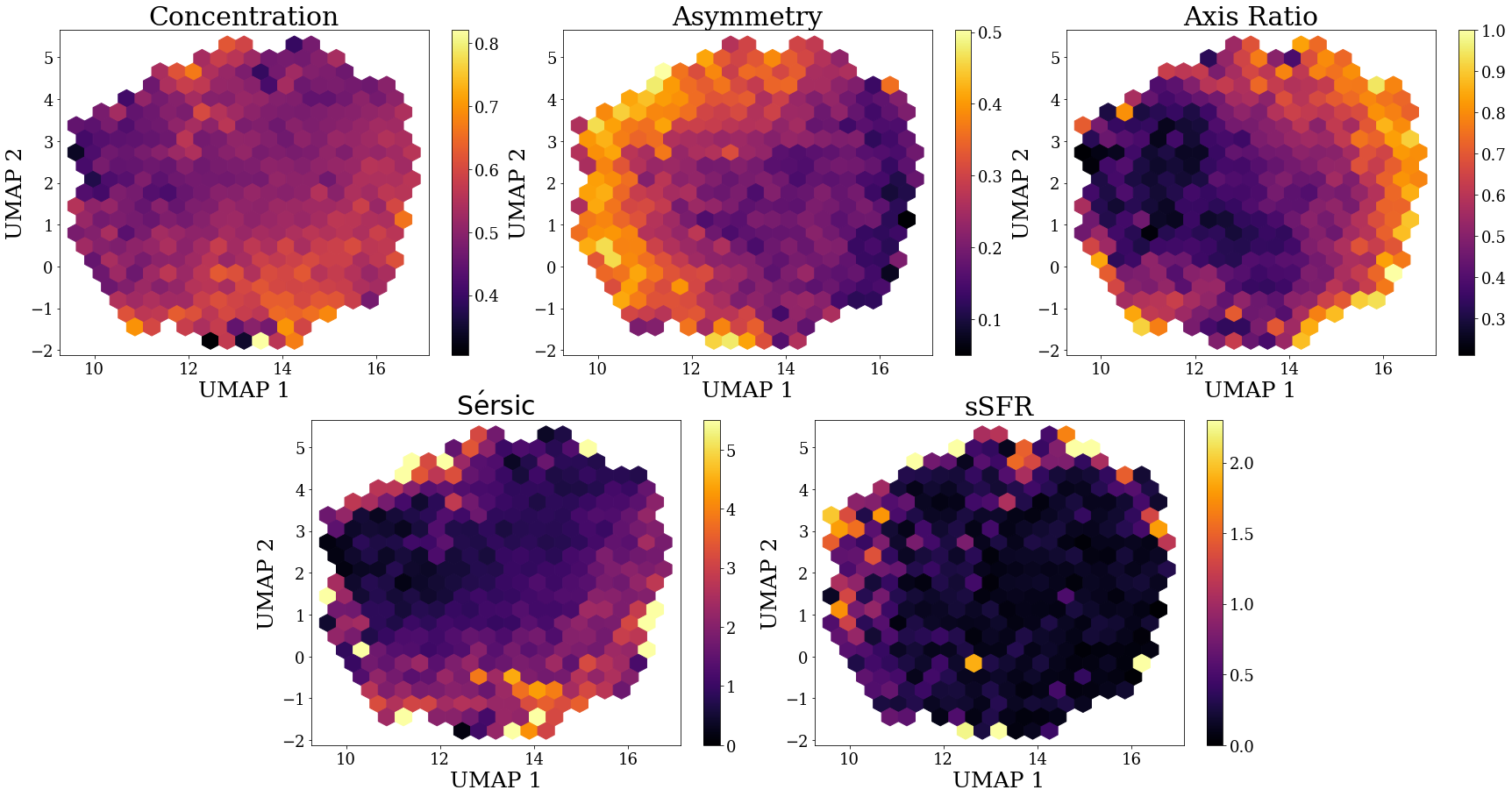}
    \caption{UMAP visualisation of our 23 latent features for our entire galaxy sample coloured by different measured morphological parameters. Each plot is coloured by the median value in each hexagonal bin in the 2D UMAP plane. From left to right shows the distribution of the concentration, asymmetry and axis ratio, S\'ersic index, and the sSFR for our galaxy sample. }
    \label{fig:UMAP}
\end{figure*}

Another method for visualising how our latent space correlates with the measured properties of the galaxies is to apply further dimensionality reduction to our 23 dimension space and represent it on a 2D plane. To do this we utilise the Uniform Manifold Approximation and Projection algorithm (UMAP, \cite{UMAP}). The UMAP algorithm seeks to produce an embedding, which is a low dimensional projection of the data, that best preserves the topological structure of the manifold. The 2D projection of our latent space is used to help visualise how the network is learning different structures in our data but it is important to remember that the coordinates of the UMAP plane  have no physical meaning and are not used for any clustering purposes in this work. The UMAP representation of our 23 dimension latent space for our galaxy sample is shown in Fig.\ref{fig:UMAP}. Each UMAP visualisation is coloured by the median value of a different measured galaxy parameter. It can be seen that as you move across the UMAP space from right to left there is a clear correlation with increasing asymmetry and decreasing concentration, S\'ersic index, and axis ratio. We can also see a trend in the sSFR with the lower sSFR galaxies residing in the lower right region of the UMAP. However, as mentioned above, the sSFR does not correlate as well as the other measurements.
From this we can see that our encoded representation space correlates with well know measured galaxy properties as expected and hence can be use to group structurally similar galaxies together which is the aim of this work. This correlation is similar to that found by \cite{vega-contrastive}. In their work they used a UMAP representation of their 1024 dimension space and showed how their space also correlates with many physical properties and how this can be beneficial when separating different types of galaxies.

\subsection{Extracted clusters}\label{sec:our_clusters}

\begin{figure*}[!htb]
    \centering
    \includegraphics[width=0.9\linewidth]{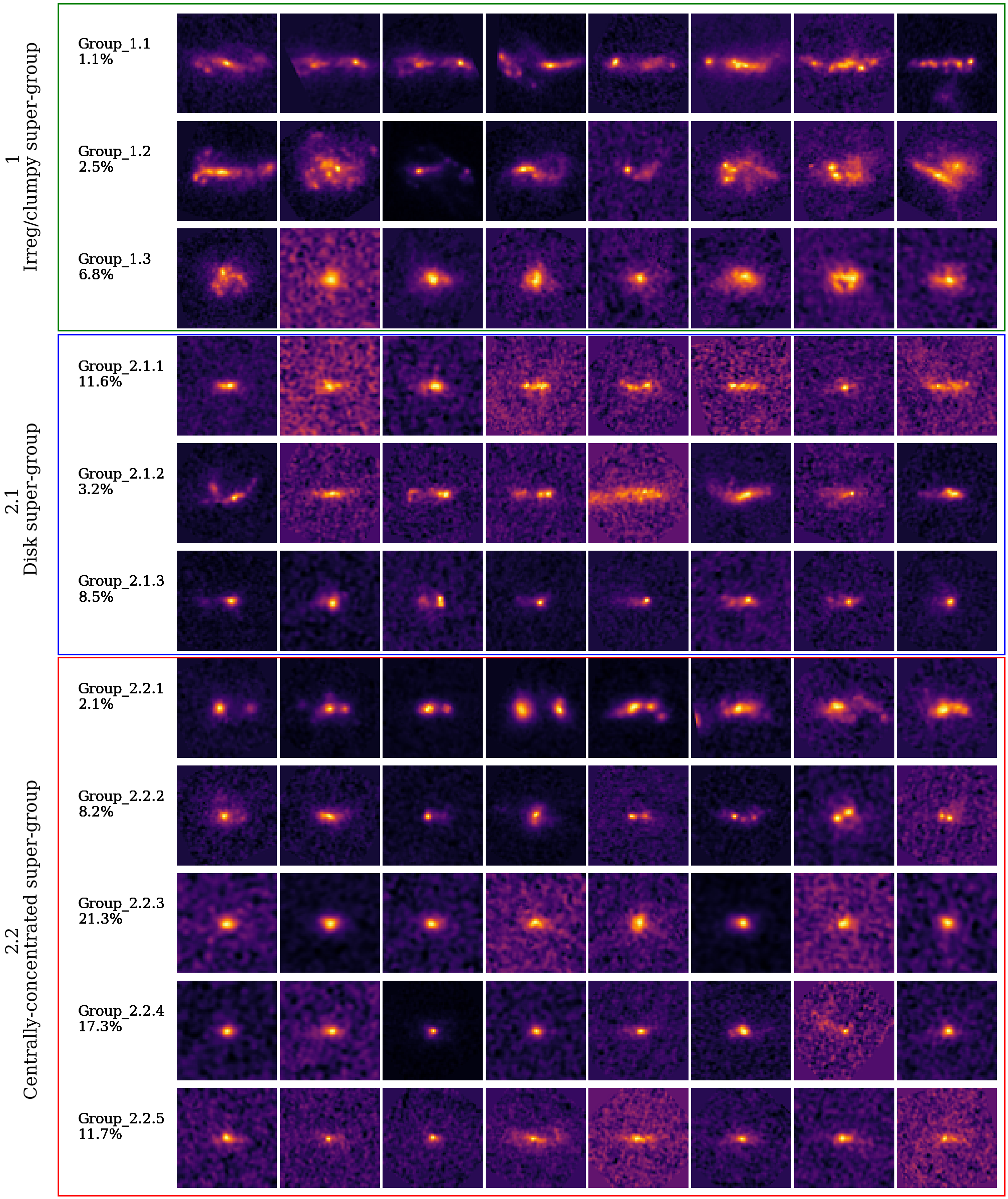}
    \caption{Examples of galaxies from each group found by the HC clustering algorithm. It can be seen that each group is visually distinct from each other. Group names indicate which parent branch they belong to from the HC tree, parent branches are labelled as super-groups along with the dominant morphology along that branch.     
}
    \label{fig:group_examples}
\end{figure*}

Running our HC clustering algorithm on our trained feature space we can investigate how the network has separated galaxies based on their morphological features and physical properties. We run the HC algorithm as explained in \S\ref{sec:clustering}. We find a total of 11 clusters after removing the clusters that extracted the lowest SNR galaxies. These low SNR galaxies accounted for 5\% of the sample and had an average signal-to-noise per pixel ($SNR_{p}$) of $SNR_{p}$$\sim$ 2 (see \cite{CASNET} for calculation of $SNR_{p}$). The median $SNR_{p}$ for the rest of our groups was $\sim 10$ for comparison. The $SNR_{p}$ ditributions for our groups can be seen in Appendix.\ref{fig:SNR_dist}.
Randomly selected cutouts of galaxies from the 11 clusters found can be seen in Fig.\ref{fig:group_examples}. Visually inspecting the galaxies within each cluster we can already see how they are visually distinct from each other. Each cluster label indicates which main/parent branch it splits from in the HC tree. This parent branch is labelled as a super-group in Fig.\ref{fig:group_examples} along with the dominant morphology of galaxies along this branch. The super-group label is not designed to be a morphological classification as each super-group is composed of galaxies with distinct structural features however, it is simply one global feature that galaxies within this group have in common. This shows how the clustering algorithm has separated out different areas in the latent space as having very separate overall shape. Super-group 2.2 for example is labelled as \textit{Centrally-concentrated} as each of the morphological classes within this group have a bright bulge like region however, group 2.2.1 is dominated by galaxy pairs or double cored so will possess structural parameters more associated with irregular type galaxies.  

Comparing our machine extracted clusters to previous works that also utilise unsupervised machine learning to extract morphological clusters, \cite{Hocking} found 200 clusters when trying to classify morphologies from the CANDELS fields. Our method results in much fewer clusters allowing for a more broad classification on the main features each galaxy image possesses. It also allows us to explore the evolution in morphology with redshift and what the main morphological features are for each group more easily than comparing 200 separate clusters. It should be noted that their work does not include as high redshifts as ours, \cite{Hocking} includes galaxies up to z$\sim$4 and also includes low-redshift galaxies which will require more information to describe their morphological features simply due to the increase in resolution. This could account for the greater number of clusters they obtain. The evolution of our machine selected clusters as a function of redshift will be explored in a follow up paper.

Whilst our machine found clusters appear to be different galaxy types by eye we want to investigate their structural and physical properties to determine if this also correlates with their apparent morphology.

\subsection{Comparison to Structural and Physical Properties}
\begin{figure*}[!htb]
    \centering
    \includegraphics[width=0.9\linewidth]{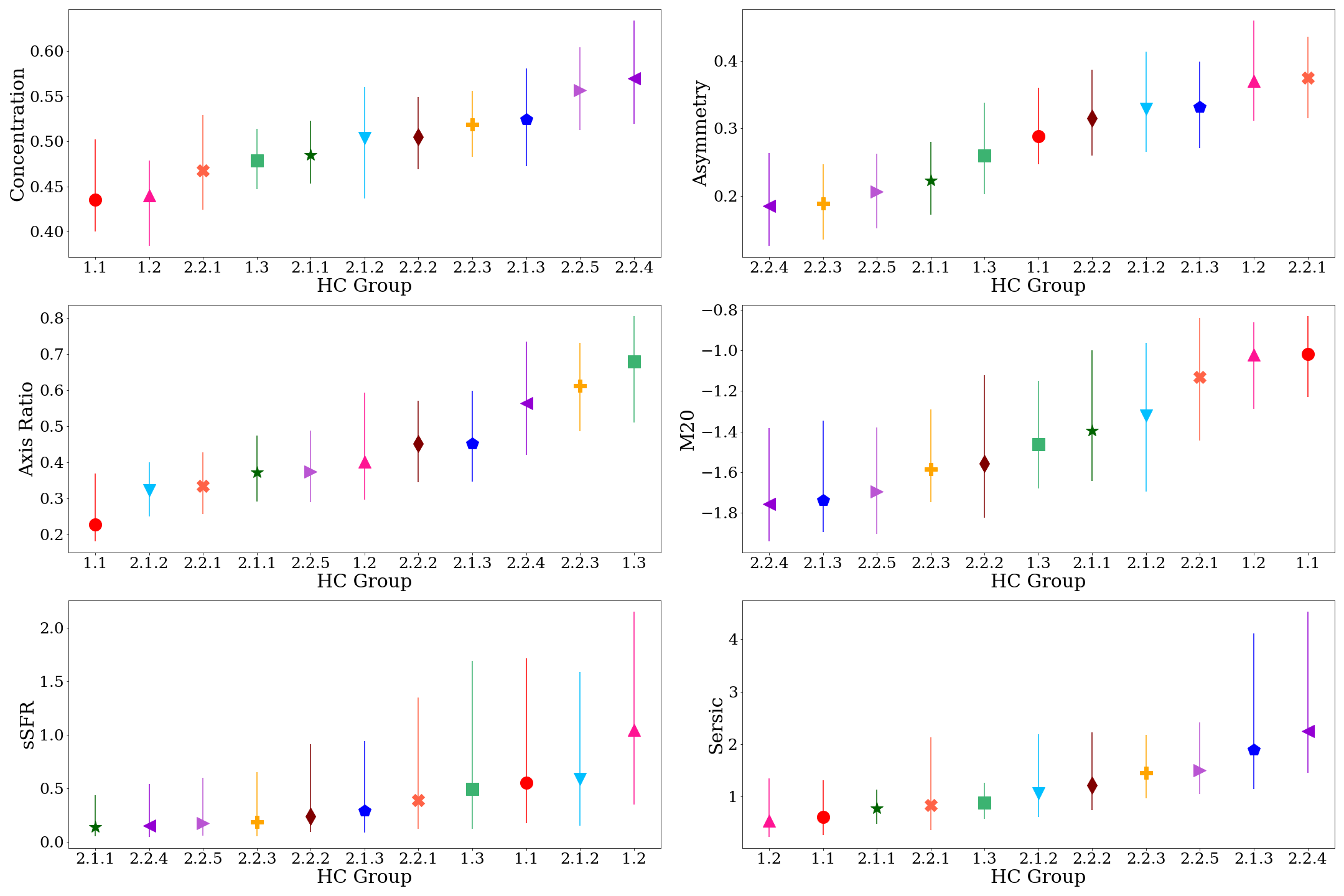}
    \caption{Distributions of the properties of the galaxies within each group determined from the hierarchical clustering algorithm. Each group is distinct from each other in both parametric and non-parametric morphology measurements and also physical properties such as the sSFR. Groups are ordered in increasing value within in measurement. Coloured points and symbols are the median and match across all figures. The error bars show 25th and 75th percentile of the distribution of values within each cluster. Note that the error on the mean is $\sim 20$ times smaller, hence the trends we see are of high significance.  
}
    \label{fig:group_params}
\end{figure*}

We use specific star formation rates (sSFRs), masses and redshifts from the original CANDELS galaxies as explained in \cite{Duncan} to ensure that the classifications are reliable, however we remeasure both the parametric and non parametric morphological parameters of each galaxy using \textsc{morfometryka} \citep{Morfometryka} as these will benefit from the higher resolution offered by JWST allowing more accurate measurements compared to those measured on the HST images. \textsc{morfometryka} uses the un-standardised images, along with the associated PSF images, to calculate these parameters. 
Investigating the distribution of these structural measurements will give a better insight into the physical properties of galaxies within each group, seen in Fig.\ref{fig:group_params}. 

Fig.\ref{fig:group_params} shows the median values of different physical and structural parameters for each of our machine defined clusters. The error bars show the 25th and 75th percentile of the distribution of values within each group. It can be seen that each of our morphological groups possess distinct physical and structural properties, even though these parameters were not used to perform the clustering. While the separation between the groups can be small considering the distribution within each group, there is still a clear transitions between the properties of the galaxies as you move between different morphological classes. Note that the error on the mean is $\sim 20$ times smaller than the interquartile range. The groups have been ordered in increasing value for each parameter to better show this smooth transition.
For example, looking at group 2.2.4, galaxies in this group  possess the highest concentration of light from all groups, most symmetric light distributions, low sSFR and high S{\'e}rsic indices agreeing with what we would classify visually as dominated by spheroidal, compact, bulge dominated systems. The same can be said for Group 1.1 and 1.2 which possess almost the opposite properties, with asymmetric light distributions, low central concentration, and the highest sSFR of the sample, agreeing with our knowledge that star forming clumps would increase the asymmetry of the galaxy's light distribution. 
We also find no separation between our groups with redshift which indicates that the network is not splitting galaxies based on any redshift effects. This reinforces the claim that our standardisation process has removed the majority of observational effects from the latent space, allowing for a more robust classification scheme. The redshift distributions of our groups are shown in Appendix.\ref{fig:redshift_dist}. 

Fig.\ref{fig:cas_sersic} shows where each cluster lies in the C-A plane as well as the distribution in S{\'e}rsic index with asymmetry. Each coloured point shows the median value for each of our machine found clusters and the error bars show the interquartile ranges of the objects within each cluster. Clusters are well separated in the C-A plane and show less, but still clear, separation with S{\'e}rsic index.
Similar results have been found by \cite{Kartaltepe_jwst} when investigating the separation of their visually defined galaxy types. In their work they showed that defining galaxy morphology by a cut in S\'ersic index would lead to a high misclassification rate between spheroids and disk galaxies due to the overlap in these mesasurments between different morphological types. This further emphasises the need for a more refined and robust classification scheme based on the morphological features of a galaxy and not on model-dependent measurements. Nevertheless Fig.\ref{fig:cas_sersic} shows a clear correlation between our clusters and the measured structural parameters of our galaxy sample. 
By using the average structural properties and visually inspecting galaxies from each group we can associate a morphological label to each which are listed in Table.\ref{tab:cluster_defs}. 
\begin{table*}[!htb]
    \centering
    \begin{tabular}{cccccccc}
        \hline
                 \textbf{Group} &  \textbf{No.samples} & \textbf{Morphological description} & \boldmath$\mathrm{C_{1/2}}$&\boldmath$\mathrm{A_{1/2}}$ & \boldmath$\mathrm{M20_{1/2}}$& \boldmath$\mathrm{sSFR_{1/2}}$ &\boldmath$\mathrm{S\Acute{e}rsic_{1/2}}$\\
             \hline
             1.1 & 58 & Edge on, multiple systems, chain, clumpy & 0.44 &0.29&-1.02&0.55& 0.60 \\ 
             1.2 & 135 & Face on multiple systems, clump clusters, mergers & 0.44 & 0.37 & -1.02&1.05&0.54\\ 
             1.3  & 365 & Clumpy disk-like, face on disk & 0.48 & 0.26 &-1.46&0.50&0.88 \\ \hline \hline
             2.1.1  & 619 & edge on disks, clumpy, single objects & 0.48 & 0.22&-1.40&0.14&0.77\\ 
             2.1.2 & 171 & Disturbed disks, tidal features  & 0.50 &0.33&-1.32&0.59&1.06\\
             2.1.3 & 455 & Tadpole galaxies, asymmetry along semi-major axis  & 0.52 &0.33&-1.74&0.29&1.90\\ \hline
             2.2.1 & 112 & Close pairs, doubles  & 0.47 &0.37&-1.13&0.39&0.84\\
             2.2.2 & 439 & Disks with tail/tidal disruption, tadpole  & 0.50 &0.32&-1.56&0.24&1.21\\
             2.2.3 & 1138 & smooth light distribution, spheroidal, elongated & 0.52&0.19&-1.58 &0.19&1.45\\
             2.2.4 & 922 & Spheroidal, bulge dominated, centrally concentrated  & 0.57 &0.18&-1.76&0.15&2.25\\
             2.2.5 & 626 & Bulge and disk component  & 0.56&0.21 &-1.70&0.17&1.50\\
    \end{tabular}
    \caption{Brief description of each group's morphology, number of samples in each cluster and median values for structural measurements. The horizontal lines show the super-group splits.}
    \label{tab:cluster_defs}
\end{table*}{}
Below we give a brief description of the main morphological features of each cluster and compare these to previous studies of high-redshift morphology.

\subsubsection{Group 1.1 - Chain galaxies}
These galaxies resemble the `chain' type morphologies first introduced by \cite{Cowie_morph}. They are very elongated structures with low axis ratios, dominated by a very elliptical shape that lacks a clear central bulge. 
\cite{Cowie_morph} stated that these extreme ellipticities argue against the possibility that these are simply galaxies viewed from edge-on and are in fact their own class of peculiar objects. 
More recently however, work by both \cite{vanderwel_disk} and \cite{Zhang_disk} suggest that there is a population of star forming galaxies that are intrinsically more elongated/prolate at higher redshifts that transition to more oblate structures later on in their evolution. \cite{Zhang_disk} also compare their results to simulations and find this to be consistent with predictions by \cite{Ceverino_sim} and \cite{Tomassetti_sim}.

\subsubsection{Group 1.2 - Clump clusters}
Group 1.2 has similar properties to our chain galaxies (group 1.1), it however possess larger axis ratios giving rise to the argument that these galaxies are simply the face-on view of chain galaxies. This argument was first put forward by \cite{Elmegreen_clumpclusters} who named these 'Clump Clusters' and stated that the distribution of axis ratios agrees with what we see for normal disk galaxies. Comparing the properties of this group with our chain galaxies (group 1.1) we see that they possess very similar distributions except for their axis ratio and asymmetry. This increase in asymmetry is consistent with the fact that these objects are face on and hence have a larger projected area resulting in a larger asymmetry. Both groups also occupy a similar redshift range and mass distribution with a median mass for group 1.1 of log(Mass[$M_{\odot}$]) = $9.60\substack{+0.26 \\ -0.38}$ and $9.64\substack{+0.50 \\ -0.37}$ for group 1.2 (where the errors denote the 25th-75th percentile range of the distribution).

\subsubsection{Group 1.3 - Clumpy disks}
These galaxies, whilst possessing a slightly clumpy morphology, are disk-dominated, with a central bulge-like region. They possess the highest axis-ratios, resembling face-on disks, but have a higher concentration than the clump clusters (group 1.2) hinting at a more evolved morphology. These systems have intermediate asymmetries possibly due to the fact they have a high sSFR which leads to some clumpy features.    
\subsubsection{Group 2.1.1 - Edge-on Disks}
Resembling edge-on disks with no central concentrated bulge region these galaxies are quite symmetric in their light distribution with intermediate concentration indicating no clear central region. They possess the lowest sSFR of our sample with low S{\'e}rsic indices. This lack of on-going star formation could indicate that these galaxies are more evolved with an outside-in formation that will perhaps form their bulge later in their evolution via means other than SF. 
\subsubsection{Group 2.1.2 - Disrupted disks}
These objects are disk dominated, single object systems that possess disrupted morphologies that could be the result of possible mergers or tidal interactions. These objects have very low axis ratios with intermediate concentrations indicating that the central galaxy is disk like, however they have very high asymmetries and $M_{20}$ values due to the tails and plumes caused by some interaction. As there is no companion visible, it is a possibility that these galaxies have been caught after a recent merger which could account for the disturbed morphology.

\begin{figure*}[!htb]
    \centering
    \includegraphics[width=0.9\linewidth]{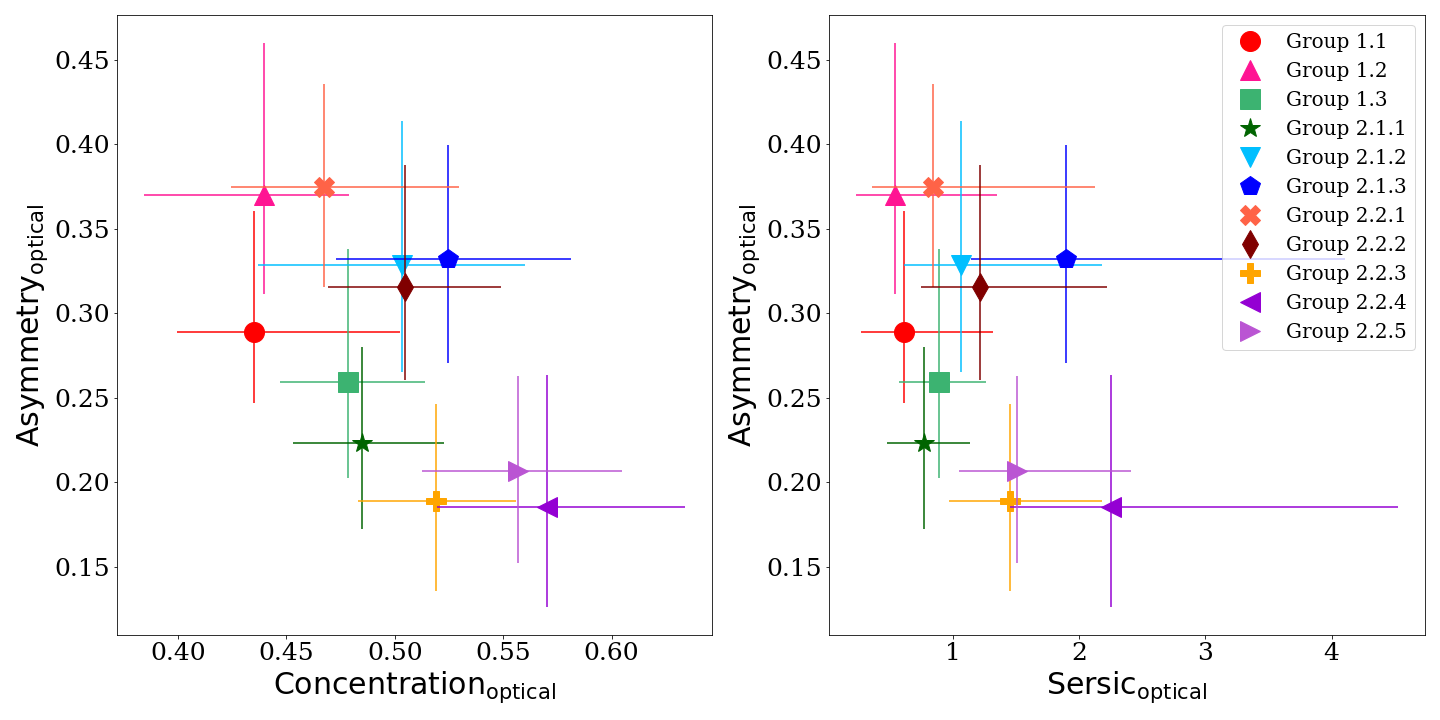}
    \caption{Distributions of the structural properties of the galaxies within each group determined from the hierarchical clustering algorithm. Each point represents one of our groups from our unsupervised machine learning clustering. Points show the median value and the error bars show the 25th and 75th percentiles of the values within each cluster. Symbols and colours as in Fig.\ref{fig:group_params}.
}
    \label{fig:cas_sersic}
\end{figure*}

\subsubsection{Group 2.1.3 - Tadpole galaxies}
Closely resembling the `Tadpole' type galaxy \citep{VanDeBergh_tadpole} with an offset nucleus with a tail resembling the shape of a tadpole. These galaxies are very asymmetric along their semi-major axis with a range of axis ratios. The origin of this structure is not fully understood, some cases could be an offset burst of star-formation, a tidal interaction, ram-pressure stripping or accretion of cosmic gas \citep{Elmegreen_tadpole}.
\subsubsection{Group 2.2.1 - Close pairs/Double clump}
Visually dominated by 2 close objects or 2 main clumps these are the most asymmetric galaxies in our sample. Similar in morphology to many previous studies such as the `Double nuclei in a common envelope' by \cite{Toomre_double} and `Double-core' galaxies in \cite{Elmegreen_clumpclusters}. 
\subsubsection{Group 2.2.2 - Tail and Tadpole galaxies}
This group possesses structural parameters very similar to our Tadpole group (Group 2.1.3) however the bright nuclei in this sample is to the left of the main object as opposed to the right with the previous group. This is due to our image standardisation process whereby each galaxy is rotated according to its position angle. This results in some tadpole galaxies being aligned $180^\circ$ opposite to others. While this removes random orientations from our network to allow meaningful features to be encoded, we cannot avoid the separation of these clusters in the feature space even though they have the same distribution of physical and structural properties. This issue will be addressed in more detail in a follow up paper.
\subsubsection{Group 2.2.3 - Elongated spheroids}
This class is dominated by symmetric, high axis ratio, low sSFR galaxies with intermediate S{\'e}rsic indices. They are spheroidal in shape but elongated, closely resembling smooth disky objects. While some may classify these objects as disks visually, their physical parameters are more in-tune with what we would expect for spheroidal galaxies. This is discussed in more detail in \S\ref{sec:vis_class}.
\subsubsection{Group 2.2.4 - Compact Spheroids}
The most concentrated and symmetric light distributions of our whole sample, these galaxies match the classic compact spheroidal type galaxies as expected from visual assessment. These galaxies are similar in properties to our elongated spheroids (group 2.2.3), they however have higher S{\'e}rsic indices, and are more centrally concentrated. These also match well with the visual classifications which are discusses in \S\ref{sec:vis_class} in more detail. They make up a total of 17.3\% of our sample.
\subsubsection{Group 2.2.5 - Bulge and Disk components}
The final class of galaxy found by our clustering technique are edge-on disks with a clear bulge component. These are separate to the other edge-on disks in our sample as they have a centrally concentrated bulge component and a clear disk component as well. The fact we see these types of galaxies at high-z is an indicator that bulge and disk formation is already in place very early on.

\subsection{Comparison to Visual Classifications}\label{sec:vis_class}
While the aim of this work is to provide a robust, non-biased approach to classifying galaxy morphology we want to compare how the networks classification system holds up against visual classifications. We have a total of 2619 visual classifications for our sample taken from \cite{leo_hubble}. While this is only $\sim$ 50\% of our sample it should still give us an indication of the average galaxy type from each HC cluster. These can be seen in Fig.\ref{fig:vis_class} (The same plot but for the super-groups is shown in Appendix.\ref{fig:vis_class_supergroup}). While there is by no means a clear correlation between the visual classifications and our clusters, this is to be expected, as the classifiers only had a limited number of labels that each galaxy had to be classified into. These are the classic spheroid, disk, peculiar and PSF labels that have been utilised in many low-z studies. There was also an option for the classifiers to choose ambiguous if they felt that no label represented the object or if they were unsure, which account for 15\% of the classifications in our sample.

Whilst we can see some correlations with the visual classifications, such as 70\% of spheroid classifications reside in clusters 2.2.3 and 2.3.4, which agree with our assessment of these clusters and their structural parameters, and groups 2.1.1 and 2.2.5 are disk dominated which also agrees with our expectations there are few other strong correlations. The aim of this work is to show that these traditional `morphological types' are not representative of the wide variety of galaxy morphology and structure we see in the high-z universe and that we need to better understand which features are important at better separating galaxies in various stages of their evolution. \cite{leo_hubble} also state in their work that there can be issues with misclassification of face on disks with spheroids and so suggest a combination of visual classifications and structural parameters would help to resolve this issue which we are combining in the feature extraction process as the network has all information about the pixel by pixel light distribution of each galaxy. In  recent work, \cite{vega-contrastive} found that a large proportion of visually classified disks perhaps lie in a region of representation space populated with spheroids. They compared their results with galaxies from TNG50 and found these regions are `occupied by objects with low stellar specific angular momentum and non-oblate structure'. Looking at our clusters we find that group 2.2.3, while having $\sim$ 45\% of the spheroid classifications, also had a large proportion of disk like classifications. This could be due, in part, to the reasons stated by \cite{vega-contrastive} above. The properties of this group more closely resemble spheroid-like galaxies with high concentration of light, low sSFR, low asymmetry and larger S{\'e}rsic indices. 

There is an argument for moving away from visual classifications all together as we now have more advanced techniques to separate galaxies based solely on their physical and measured features alone. With the increase in the amount of data being collected every day, visually inspecting each galaxy is also inconceivable. Measuring the properties of each galaxy accurately however, requires detailed spectral analysis which is time consuming and again would be inconceivable for the amount of data now available. Performing feature extraction on galaxy images unifies these two methods together and can be trained on a small sub-sample of images with detailed measurements and then applied on a much larger, un-labelled dataset, ideal for the future of galaxy surveys and galaxy evolution studies.

\begin{figure}[!htb]
    \centering
    \includegraphics[width=0.9\linewidth]{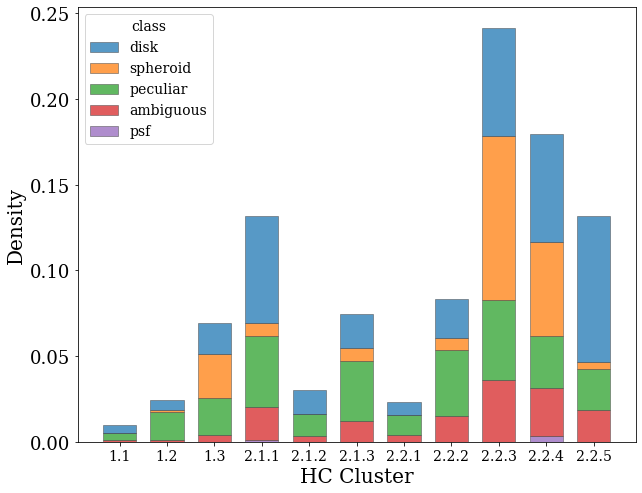}
    \caption{Barplot of visual classifications from \cite{leo_hubble} within each of our HC found clusters. We have a visual classification for $\sim$ 50\% of our sample.  
}
    \label{fig:vis_class}
\end{figure}

\subsection{Impact of noise} \label{sec:noise}
\begin{figure*}
    \centering
    \includegraphics[width=\linewidth]{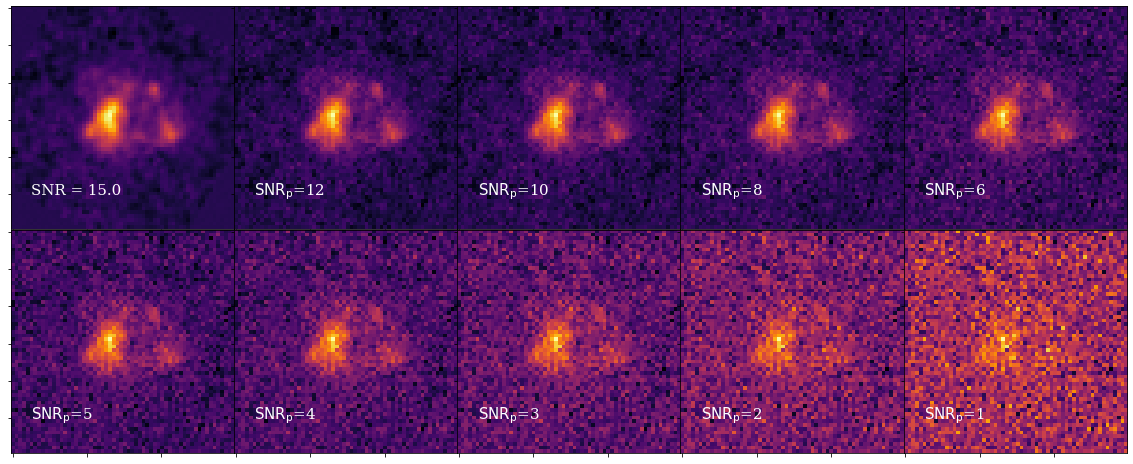}
    \caption{An example of our simulated noisy galaxy images. In the top left panel we show the original JWST cutout followed by the same galaxy at different simulated $\mathrm{SNR_p}$ levels. }
    \label{fig:snr_images}
\end{figure*}

\begin{figure}
    \centering
    \includegraphics[width=\linewidth]{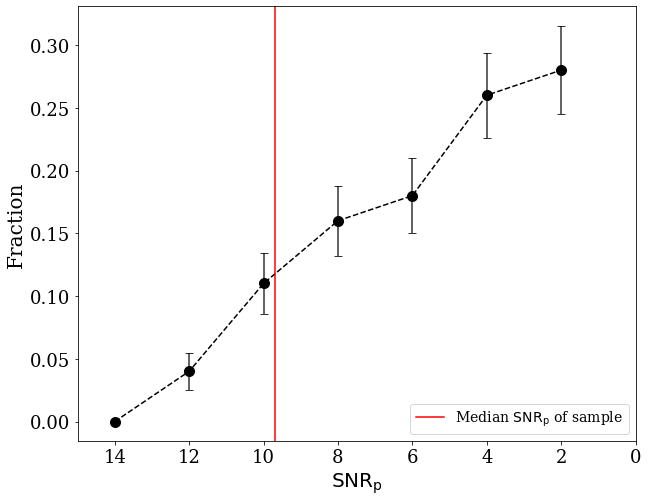}
    \caption{Fraction of galaxies that change classification in our machine-defined clusters vs the $\mathrm{SNR_p}$ of the galaxy. The errors are given as 1$\sigma$ of the binomial distribution. The red vertical line shows the median $\mathrm{SNR_P}$ of our full sample.  }
    \label{fig:SNR_frac}
\end{figure}

One common issue with visual classifications and both parametric and non-parametric measurements is that they are sensitive to noise \citep{VDB_spiral,Morfometryka, Gullberg_sersicnoise}. This is another motivation behind utilising UML to extract different morphological features, and why we optimise our latent space to focus on encoding the galaxy light and not the background noise (see \S\ref{sec:dimension}). While the network is performing well at separating out distinct morphological clusters of galaxies, we want to evaluate how robust these clusters are to the signal-to-noise level of our images. To test this we select a a sample of 15 high signal-to-noise ($14 < \mathrm{SNR_p} < 20$) galaxies from each machine-defined cluster and simulate each galaxy at various $\mathrm{SNR_p}$ levels from 12 down to 2 (for details of how these images are simulated see \citealt{CASNET}). An example of the simulated noisy images can be seen in Fig.\ref{fig:snr_images}. We assume for this test that these high $\mathrm{SNR_p}$ images are in their `true' clusters, as the majority of their features should be detectable by the network.
We then run the simulated images through the trained network and the HC algorithm to check what fraction of galaxies remain in the same morphological cluster. The results are plotted in Fig.\ref{fig:SNR_frac}. The black points show the fraction of galaxies where their classification changed and the error bars show 1$\sigma$ of the binomial distribution. The first point is plotted at $\mathrm{SNR_p}$ of 14, and is at zero as we are assuming these galaxies to be in the correct cluster. The median $\mathrm{SNR_p}$ of our entire galaxy sample is shown by the red vertical line. It can be seen that at the median $\mathrm{SNR_p}$ of our sample, 87-91\% of our galaxies remain in their original machine-defined cluster. This means that per group we would expect $\sim 1\%$ to change class at this $\mathrm{SNR_p}$ level. As the $\mathrm{SNR_p}$ in the images continues to decrease we see an increase in the fraction of galaxies that change classification. Groups 2.1.1 and 2.1.2 had the largest fraction of galaxies change class, while group 2.2.3 remained the most consistent. This is to be expected as when the noise in an image increases you start to lose fainter features from the images rapidly. This means that, for example, galaxies defined as having tidal features in our classifications (group 2.1.2) would no longer be differentiable from the background noise by the network, and end up in a different morphological category defined by the remaining detectable features. This is not something that can be fully addressed in any classification scheme as this would also mean these features would no longer be picked up in a visual classification, parametric, or non-parametric measurement system. A good example of this can be seen in Fig.\ref{fig:snr_images}, where the fainter structures in the galaxy become indistinguishable from the background noise around $\mathrm{SNR_p}$=5.
From this we can argue that the different morphological clusters found by our method are robust to various levels of noise however, below a $\mathrm{SNR_p}< 6$ there is less certainty of the morphological classification from this method. For reference, 82\% of our sample has a $\mathrm{SNR_p} > 6$ so we can be confident in the majority of our morphological classifications.

\subsection{Evolution of Massive Galaxies}

\begin{figure*}[htb]
    \centering
    \includegraphics[width=0.95\linewidth]{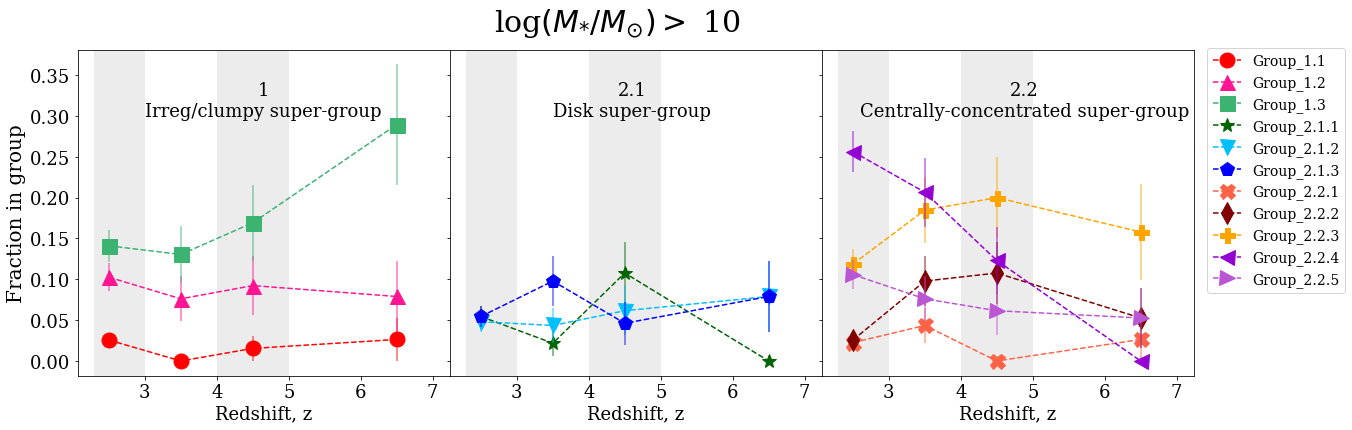}
    \caption{Evolution of the most massive galaxies within each morphological cluster with redshift. The errors are given as $1\sigma$ of the binomial distribution. Colours and symbols as in Fig.\ref{fig:group_params} and Fig.\ref{fig:cas_sersic}. The groups are divided by their HC super-group for ease of visualisation and comparison. The shaded regions show the bin widths for each point. The last bin includes all galaxies above $z > 5$. See text for details.
}
    \label{fig:highmass_evo}
\end{figure*}

As this work covers a large redshift range we can start to explore the evolution of the different clusters with redshift. To investigate these trends we plot the evolution in the fraction of galaxies in each cluster splitting into 4 bins: $2<z<3$, $3<z<4$, $4<z<5$, and $z > 5$. 
In order to ensure any conclusions drawn are not affected by incompleteness we look at the most massive galaxies (log$M_{*}/M_{\odot} >$ 10) within our sample across the whole redshift range. This can be seen in Fig.\ref{fig:highmass_evo}. The groups are split by their HC super-group for ease of visualisation and so it is easier to compare different groups. We have a total of 507 galaxies above this mass and so at the highest redshift bins we can be affected by small sample statistics which is reflected in the errors at these values. Consequently the last bin includes all galaxies above $z > 5$ as there are not enough objects at this redshift to split this further. For this reason there is less certainty of the evolution after $z > 5$ and will need larger samples to study this in depth which will be possible in the near future with more data from JWST.  Whilst there are not many galaxies at these masses we can still see some clear trends especially with group 2.2.4 (our spheroid dominated class) decreasing with redshift which agrees with results found in other works using similar data from JWST. In \cite{Huertas-company_jwst} they found a decrease in early type/bulge dominated systems with redshift as did \cite{Kartaltepe_jwst} who found a decrease in spheroid only classified galaxies at the high masses (although they note this could be due to faint features being missed or small number statistics). When we reduce our mass cut to match the analysis in \cite{Kartaltepe_jwst} of $ > 10^{9}M_{\odot}$ and investigate our 2 spheroid dominated classes (2.2.3 and 2.2.4) we find that it decreases from $\sim$ 38\% at z = 3-4 to $\sim$ 32\% at z $>$ 5 similar to their findings that their spheroid only class falls from $\sim$ 42\% at z = 3 to $\sim$ 30-40\% at z $>$ 5.  These results are fairly similar even though our sample size is more than double the sample used by \cite{Kartaltepe_jwst} at this mass range. 
We also see an increase in groups 1.3 and 2.1.2, which are various types of disk dominated galaxies in our classification system but are distinct from each other in terms of being face-on disks and  disturbed disks respectively. This result agrees with many works that find disks to be dominant or already in place at high-z \citep{leo_hubble, leo_panic, Huertas-company_jwst}. We also see a growth in group 2.2.5 which are galaxies with a distinct bulge and disk component indicating that the process of forming a bulge with a disk component is already in place very early on agreeing with some of the main findings from \cite{Huertas-company_jwst}. 
We acknowledge that these findings are subject to small sample statistics and can be further explored and solidified with future surveys and more data especially at the high-z end but it is a good sign that even with the variety of morphologies that are found by exploring the networks extracted features agree with various other works using very different methods.
We will study the evolution of these galaxies with redshift in more depth in a follow up paper.

\subsection{Morphology of High sSFR Galaxies}\label{sec:sSFR}
\begin{figure}[!htb]
    \centering
    \includegraphics[width=0.9\linewidth]{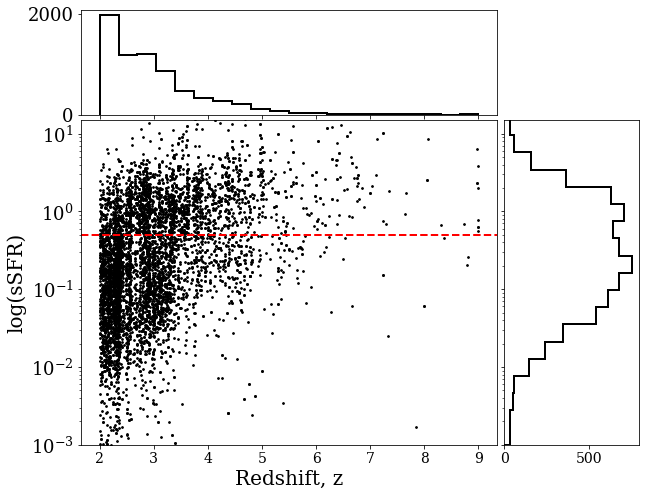}
    \caption{Distribution of the sSFR of galaxies within our matched sample vs redshift. Red line shows sSFR cut at $\log(\mathrm{sSFR}) > 0.5$. 
}
    \label{fig:sSFR_cut}
\end{figure}

We analyse the morphology of high star forming galaxies within our machine found clusters by making a cut at log$(sSFR) > 0.5$ as this is representative of our galaxy population in the high-z regime. This cut can be seen in Fig.\ref{fig:sSFR_cut}. While we are most likely missing a lot of the fainter and lower mass galaxies at high-z, we expect to be more complete for the high sSFR galaxies as this high star-formation rate is what allows us to detect even the most distant galaxies. 

We find that 4 out of our 11 machine-defined clusters have a median sSFR above this cut and so we can begin to investigate the variation in morphology seen for these high star-forming systems. The 4 groups are the Chain galaxies (group 1.1), the Clump Clusters (group 1.2), Clump Disks (group 1.3), and Disrupted Disks (group 2.1.2). Both the Chain and Clump Cluster groups possess galaxies with the highest sSFRs in our sample which is to be expected as it is believed the bright knots dominating their morphology are areas undergoing intense star formation \citep{Cowie_morph, Elmegreen_clumpclusters}. These star-forming clumps are picked up easily by structural parameters as these groups also possess the lowest concentration of light and S{\'e}rsic indices, and possess high asymmetry values. On top of that they are also the two highest $\mathrm{M_{20}}$ groups, showing that these systems could be picked up easily using traditional methods. The Clump Disk galaxy type in our classification scheme appear visually to be slightly messy, containing some clumpy morphology, which could indicate they are undergoing intense star-formation, however looking at the non-parametric measurements alone we find this group to have intermediate CAS and $\mathrm{M_{20}}$ values and so could be missed if selection cuts were to be made using these structural measurement systems. They are however found by our classification system, showing again that our network is able to pick up on features that are missed with current measurements. The final group we look at is the Disturbed Disks which, like our Clump Disk group, are asymmetric but possess intermediate concentration of light corresponding to a bulge component indicating that these systems are more evolved than the Chain and Clump cluster systems. These systems do not possess the classic star-forming clumpy morphology but instead have an asymmetry in the form of a tail or disturbed region. As these are all individual systems with no clear neighbour, it could be argued these galaxies have recently undergone a merger which could account for both the disturbed morphology, and corresponding increased star-formation.
From these groups we can already see how diverse and varied the morphologies of these high sSFR galaxies are and why it is important to understand the processes that lead to this diverse structure if we want to better constrain their formation and evolution.
We acknowledge that the classifications presented in this work do not allow us to confirm the formation history of these systems, however the ability to separate out these high sSFR systems with different morphologies for future, more detailed analysis and observations allows for an unbiased and robust selection process. Again it should be noted that more detailed analysis of the evolution of these morphological groups with redshift will be carried out in a follow up paper.

\section{Summary}\label{sec:summary}
We present our work utilising unsupervised machine learning to perform feature extraction on high-redshift galaxies imaged with JWST. We apply a hierarchical clustering algorithm to extract separate, self-similar, morphological classes of galaxies; resulting in a robust, more meaningful classification system of these objects. This is the highest redshift sample to date using this technique. 
Applying our method to optical rest-frame galaxy images imaged with NIRCam on JWST we find 11 separate morphological clusters that possess different morphological features, physical properties and structural measurements e.g. sSFR, \CAS-$M_{20}$ parameters, S{\'e}rsic indices and axis ratios. Our resulting clusters are devoid of human biases and would not be as well separated if classified with traditional nomenclature. We compare our findings with visual classifications and find that only spheroids are well separated in this traditional classification system and that disks and peculiar type galaxies need much more detailed descriptions.  
We improve upon previous studies using similar methods in multiple ways;
\begin{itemize}
    \item We remove the observational biases imposed on the images by standardising our sample before performing any feature extraction. This allows the network to focus on the morphological features of the target galaxy without wasting information encoding the position angle of the galaxy, background sources, and noise, leading to a more physically meaningful feature space. 
    \item Our method results in many fewer, well separated, morphological classes that can be investigated in detail which is not possible in some previous studies when hundreds of clusters are extracted. 
    \item We have a relatively small feature space that can be investigated and linked to individual structural properties leading to clusters that are well separated in both parametric and non-parametric parameters. 
    \item We explore the highest redshift sample to date utilising unsupervised machine learning. Thanks to JWST we also have access to rest-frame optical imaging across all redshifts so our classification system is not biased by variations between UV and optical morphologies. 
\end{itemize}

\noindent Due to the wide redshift range covered in this work we have access to a wide span of early history of galaxy formation which allows us to investigate various trends with cosmic time. Our main findings are summarised below.
\begin{itemize}
    \item We find that there is a wide variety of galaxy morphology already in place at high-z. In total we find 11 distinct morphological types for our sample. 
    \item We confirm that our unsupervised machine-defined clusters support work to construct a visual classification scheme suitable for high-z, while side-stepping the issue of applying pre-defined categories to new observational regimes. 
    \item We find a decrease in concentrated spheroidal type galaxies with redshift as found by others, and find that disk-like galaxies dominate at high-z, though these are typically clumpy and/or disturbed in morphology. 
    \item Unsupervised methods allow us to establish which morphological features are important and have an impact on the physical properties of the galaxies themselves. The resulting extracted features will provide a more detailed and better suited classification system.
\end{itemize}
As mentioned in the paper, we plan to carry out more detailed studies with redshift evolution and link the morphological classes found in this work to the low redshift universe. With the accumulation of data from JWST we expect our view of the distant universe to continue to expand and improve with more and more observations and detailed analysis, all of which will help improve galaxy evolution studies such as the work carried out in this paper. Such approaches to galaxy morphology classification are also required to handle the amount of data expected with the future of JWST and similar surveys. 

\section{Acknowledgements}
The authors would like to thank the University of Nottingham for providing the computational infrastructure needed to produce the networks used in this paper. We also thank Phil Parry for timely help with this infrastructure that made this work possible.  CT and TH acknowledges funding from the Science and Technology Facilities Council (STFC). We acknowledge support from the ERC Advanced Investigator Grant EPOCHS (788113). We thank the CEERS team for making their data open to the public, which we have made use of within this work. We thank the anonymous referee for their thorough review, which helped to significantly improve the presentation of this work.
This work is based on observations made with the NASA/ESA \textit{Hubble Space Telescope} (HST) and NASA/ESA/CSA \textit{James Webb Space Telescope} (JWST) obtained from the \texttt{Mikulski Archive for Space Telescopes} (\texttt{MAST}) at the \textit{Space Telescope Science Institute} (STScI), which is operated by the Association of Universities for Research in Astronomy, Inc., under NASA contract NAS 5-03127 for JWST, and NAS 5–26555 for HST. The specific observations analysed can be accessed via \dataset[10.17909/z7p0-8481]{https://doi.org/10.17909/z7p0-8481}.

\bibliography{ref}{}
\bibliographystyle{aasjournal}

\appendix
\setcounter{figure}{0}                       
\renewcommand\thefigure{A\arabic{figure}} 

\section{Dimensionality of the latent space}

\begin{figure}[h!]
    \centering
    {{\includegraphics[width=8cm]{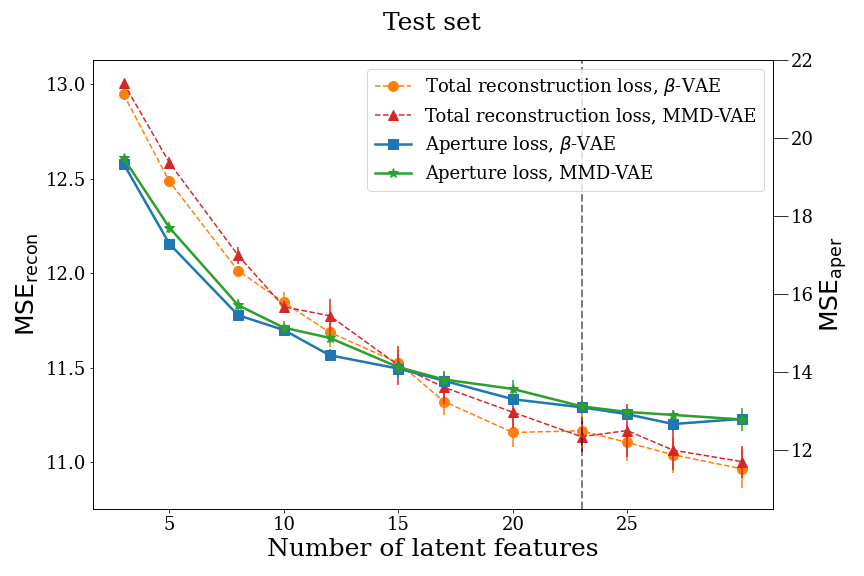} }}%
    \qquad
{{\includegraphics[width=8cm]{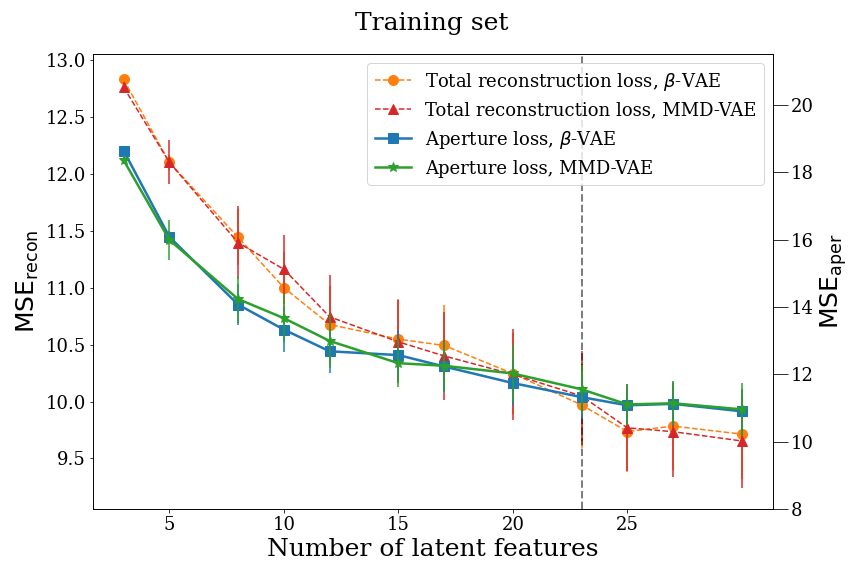} }}%
    \caption{Variation in the total reconstruction loss and the reconstruction loss within an aperture (see  \S\ref{sec:dimension} for details) for the test and training set (same as in Fig.\ref{fig:latent_opt}). olid points show the average MSE for the last 50 epoches and error bars show 1$\sigma$ of the variation. Dashed line shows the chosen dimensionality of the latent space for our analysis.}%
    \label{fig:network_opt}%
\end{figure}

In Fig.\ref{fig:network_opt} we show the variation in the reconstruction of both the training and test sets vs the dimensionality of the latent space. Solid points show the average MSE for the last 50 epoches (our patience value, see text for details) and error bars show 1$\sigma$ of the variation.
As with the validation set (see Fig.\ref{fig:latent_opt}) the reconstruction loss will continue to improve with more latent features however the aperture loss plateaus off. This indicates that the network is using the extra dimensions to encode some background noise in the images which we do not care about.  

\section{visualising the latent space}
\begin{figure}[!htb]
    \centering
    \includegraphics[width=0.9\linewidth]{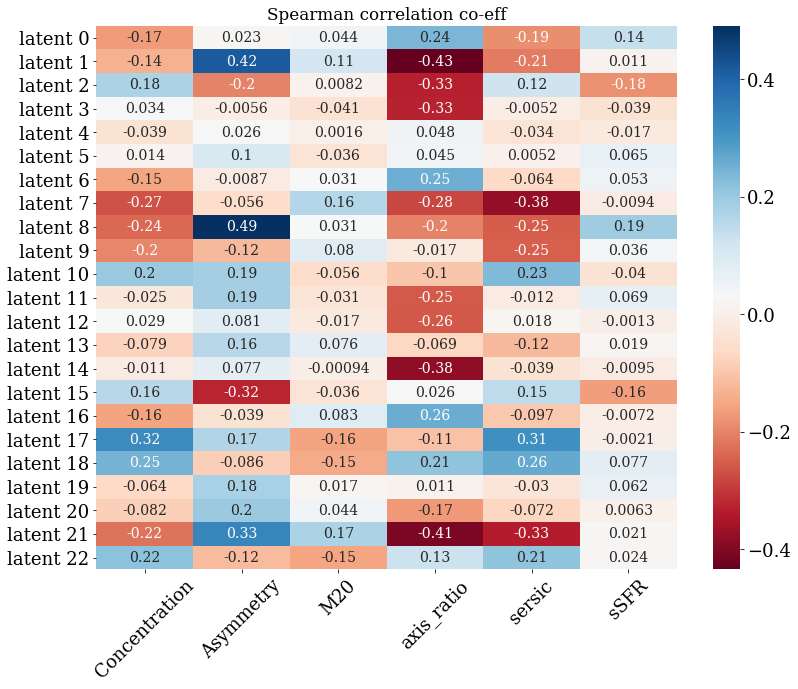}
    \caption{Distribution of the Spearman rank correlation coefficients for each latent dimension in our trained MMD-VAE and each measured galaxy property.}
    \label{fig:spearman}
\end{figure}
In Fig.\ref{fig:spearman} we show the full Spearman correlation matrix for each of our 23 latent dimensions and the measured galaxy properties. In Table.\ref{tab:co-effs} we list the highest correlated latent variable for each of the galaxy properties. Note that each latent variable is sensitive to a particular combination of particular attributes, for example latent feature 1 correlates with both asymmetry and axis ratio but not with concentration or M20. We can see that it is not a one to one correlation between each latent dimension and each physical parameter indicating that the network is using multiple latent features to encode each physical property. The fact that there are combinations of correlations between features and latent dimensions and not just random noise shows that the network is learning the different features of each galaxy, thus allowing the clustering of the latent space to provide different morphological types of galaxies.

\begin{figure}
    \centering
    \includegraphics[width=0.8\linewidth]{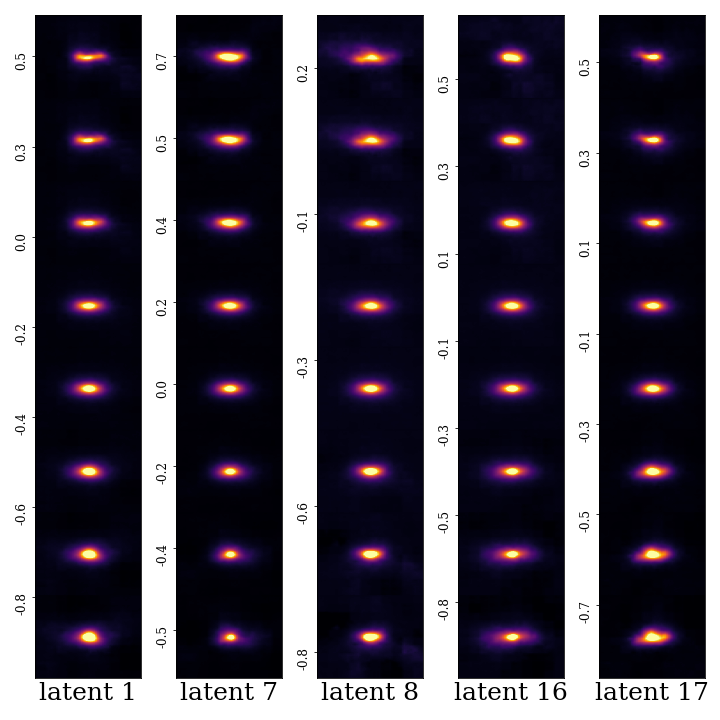}
    \caption{Generated images from the trained latent space. Here we show examples of generated galaxy images along different dimensions in the latent space to show the smooth transition between morphology. The latent dimensions shown are those with the highest correlation to measured properties (see Fig.\ref{fig:latent_vs_feat} for more detail.) We calculate the mean of each feature and linearly interpolate between $\pm 2\sigma$ whilst keeping all other latent features constant to show the morphology encoded by each dimension.
    }
    \label{fig:component_plot}
\end{figure}

In Fig.\ref{fig:component_plot} we show generated images from our trained latent space. Plotted are the latent features that had the highest Spearman correlation coefficients with measured physical and structural properties. These are not real galaxy images but images generated from the encoded space learned by our network. We generate these images by calculating the mean and range of each latent feature and then interpolating along each feature generating a new image with the network each time, while the other latent features are keep constant. This shows that our network has learnt a smooth feature space (which is expected with a VAE) as the morphology transitions smoothly along, and between, each latent feature. We can also see correlations with known morphology. Visually we can see how the first latent feature correlates well with axis ratio, while feature 7 correlates well with S\'ersic index. Another important morphological feature is the bulge to disk ratio which is well encoded in feature 16.

\section{Correlation of latent space with non-physical properties }
\begin{figure}[!htb]
    \centering
    \includegraphics[width=0.8\linewidth]{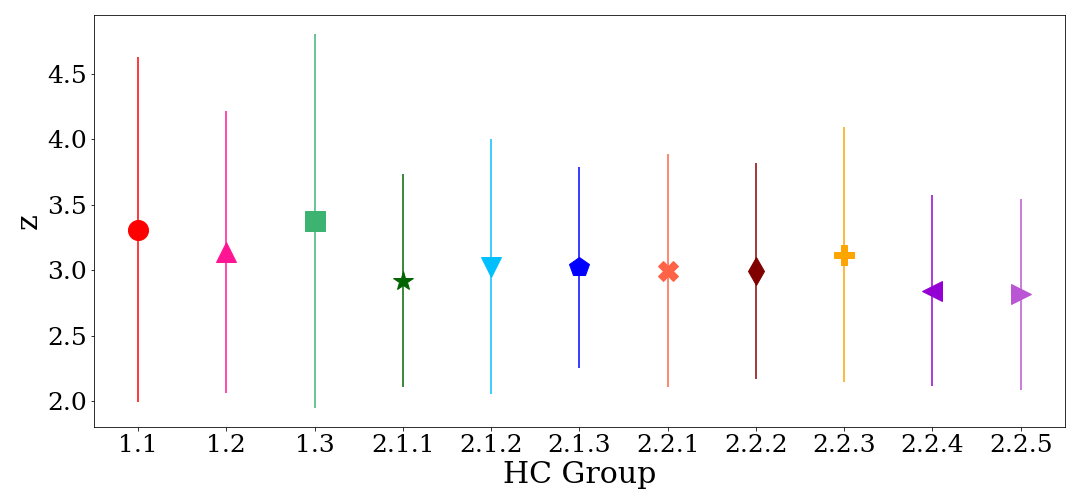}
    \caption{Distributions of the redshifts of galaxies within each morphological cluster. It can be seen that the network does not split groups based on their redshift which further supports the claim that our standardisation approach successfully removes many observational biases such as apparent size/ apparent brightness. This also allows further study on the evolution of galaxies within each morphological class with cosmic time. Coloured points and symbols are the same as the other figures in this work. Solid points show the mean and error bars show the standard deviation within each group.}
    \label{fig:redshift_dist}
\end{figure}

\begin{figure}[!htb]
    \centering
    \includegraphics[width=0.8\linewidth]{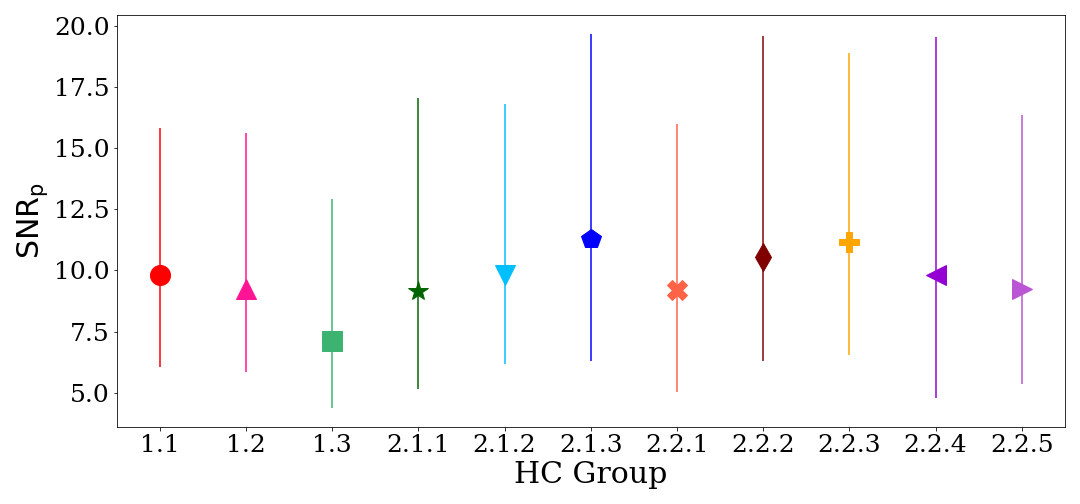}
    \caption{Distributions of the $SNR_{p}$ of galaxies within each morphological cluster. It can be seen that the network does not split groups based on their $SNR_{p}$. Coloured points and symbols are the same as the other figures in this work. Solid points show the mean and error bars show the standard deviation within each group.}
    \label{fig:SNR_dist}
\end{figure}

In Fig.\ref{fig:redshift_dist} we show the distribution of the redshifts within each morphological class. It can be seen that there is no significant difference between any of the groups and that they all span similar ranges indicating that our network is not splitting based on redshift effects or different observational bands. This further supports the power of our standardisation process at removing observational effects from our images such as apparent size and brightness as otherwise the network would have used these features to separate the groups. It can be seen that groups in the 1st super-group have slightly larger errors than the other groups but this is due to the small number of galaxies in each of these groups as they are the most irregular of our sample. The same can be seen for the $SNR_{p}$ of each group in Fig.\ref{fig:SNR_dist}. We see that the network is not using the noise in an image to split the groups. This is important as we have taken care to try to prevent the network from learning noise in an image as a feature (see \S\ref{sec:dimension}). 

\section{Comparison to visual classifications}
\begin{figure}[!htb]
    \centering
    \includegraphics[width=0.8\linewidth]{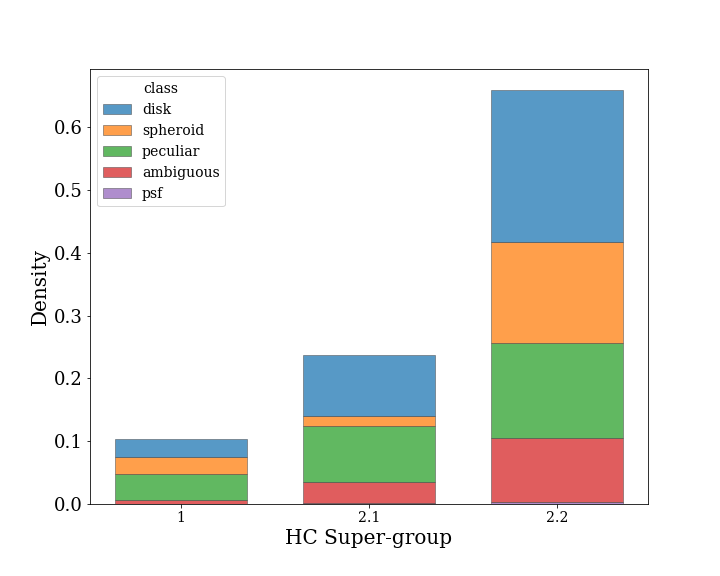}
    \caption{Barplot of visual classifications from \cite{leo_hubble} within each of our HC found super-groups. We have a visual classification for $\sim$ 50\% of our sample. }
    \label{fig:vis_class_supergroup}
\end{figure}

In Fig.\ref{fig:vis_class_supergroup} we show the comparison between our HC super-groups and the visual classifications from \cite{leo_hubble}. As it can be seen there are no strong correlations except for the centrally concentrated super-group (2.2) possessing most of the spheroid classifications. This further supports the argument that the visual classification systems used at low redshift are not suitable for the high redshift regime. 

\end{document}